\documentclass[a4paper,11pt]{article}

\usepackage[dvips]{graphicx}
\usepackage{amsfonts, color, float, amsmath}
\usepackage{amssymb}\usepackage{subfigure}
\usepackage{url}
\usepackage{array}
\usepackage[normalem]{ulem}
\usepackage{slashed}
\usepackage[dvipsnames]{xcolor}

\usepackage[small,bf,sf]{caption}
\usepackage{microtype}
\usepackage{cite}
\usepackage{authblk}

\usepackage{xspace}

\usepackage[bottom]{footmisc}
\usepackage{booktabs}
\usepackage[linktoc=all]{hyperref}

\usepackage{multirow}

\usepackage{colortbl}
\definecolor{myblue}{rgb}{0.8,0.85,1}
\definecolor{light-gray}{gray}{0.95}

\hypersetup{colorlinks=true,
citecolor=ForestGreen,
anchorcolor=CornflowerBlue,
linkcolor=NavyBlue,
urlcolor=Mahogany,
linkbordercolor={1 1 0}}

\allowdisplaybreaks

\newcommand{\MSb}{{\overline{\rm MS}}}

\newcommand{\sand}[3]{\langle#1|#2|#3 \rangle}

\newcommand{\epem}{e^+e^-\to {\rm (hadrons)}}

\newcommand{\Lb}{large-$\beta_0$\xspace}
\newcommand{\mbar}{\overline{m}}
\newcommand{\df}{{\rm d}}

\def\beq {\begin{equation}}
\def\eeq {\end{equation}}
\def\bea {\begin{eqnarray}}
\def\eea {\end{eqnarray}}
\def\G{\Gamma}

\def\nn {\nonumber}

\def\epem{e^+e^-\to(\mbox{hadrons})}

\AtBeginDocument{}

\title{\LARGE {\bf \sffamily \boldmath Small-momentum expansion of heavy-quark correlators in the large-$\beta_0$ limit and $\alpha_s$ extractions}}

\author[1,2]{Diogo Boito}
\author[3,4]{Vicent Mateu}
\author[1]{Marcus\,V. Rodrigues\thanks{marcus.gonzalez.rodrigues@usp.br (corresponding author)}\vspace{0.3cm}}

\affil[1]{\it Instituto de F\'isica de S\~ao Carlos, Universidade de S\~ao Paulo, CP 369, 13560-970, S\~ao
Carlos, SP, Brazil\vspace{0.3cm}}

\affil[2]{\it University of Vienna, Faculty of Physics, Boltzmanngasse 5, A-1090 Wien, Austria\vspace{0.3cm}}

\affil[3]{\it Departamento de F\'isica Fundamental e IUFFyM,
Universidad de Salamanca, E-37008 Salamanca, Spain\vspace{0.3cm}}

\affil[4]{\it Instituto de F\'isica Te\'orica UAM-CSIC, E-28049 Madrid, Spain\vspace{0.3cm}}

\date{}

\begin{document}
\begin{flushright}
{\small
UWThPh 2021-6\\
IFT-UAM/CSIC-21-70 \\
\today}
\end{flushright}

\vspace*{1.cm}
\begingroup
\let\newpage\relax
\maketitle
\endgroup
\date{}

\vspace*{0cm}
\begin{abstract}
\noindent
We calculate the small-momentum expansion of vector, axial-vector, scalar, and pseudo-scalar heavy-quark current correlators in the large-$\beta_0$
limit of QCD, extending the analysis of Grozin and Sturm beyond the vector current. Our results are used to study the
higher-order behaviour of dimensionless ratios of vector and pseudo-scalar moments used for the precise extraction of the strong coupling,
$\alpha_s$, from relativistic quarkonium sum rules and lattice data, respectively. We show that these ratios benefit from a partial cancellation of
the leading renormalon singularities. Our results can guide the design of combinations of moments with improved perturbative behaviour.
\end{abstract}

\thispagestyle{empty}

\newpage
\tableofcontents

\setcounter{page}{1}
\section{Introduction}
In the absence of direct observation of new physics at the LHC, precision physics remains a crucial tool to search for phenomena beyond the Standard Model. With the recent developments in multi-loop calculations in Quantum Chromodynamics (QCD), theoretical uncertainties in several key observables are now dominated by the errors on the fundamental QCD parameters, namely the quark masses and the strong coupling, $\alpha_s$.
It is therefore essential to achieve an excellent control over these quantities.
With the forthcoming $e^+e^-$ facilities that should aim at Higgs and top-quark mass precise measurements, a good control of $\alpha_s$ as well as the charm-, bottom-, and top-quark masses will remain central for the determination of constraints on the Standard Model and searches for physics beyond it.

One of the most frequently used tools for the precise extraction of the charm- and bottom-quark masses are QCD sum rules~\cite{Shifman:1978bx,Shifman:1978by}, where theory predictions are related
to measurements of the inclusive hadronic $e^+e^-$ cross-section through weighted integrals over the $R_{q\bar q}(s)$ ratio. The inverse moments defined as
\begin{equation}
M_{q,n}^{V} = \int_{s_{\rm th}}^{\infty} \frac{\df s}{s^{n+1}}R_{q\bar q}(s)\,, \label{eq:MnV}
\end{equation}
with $n\geq 1$ and $q=c,\,b$, are highly sensitive to the heavy quark mass and play a central role in this program. With the use of analyticity and unitarity constraints,
these moments can be related to the coefficients of the small-momentum (below threshold) expansion of the quark vector-current correlator.
These coefficients, in turn, can be calculated reliably in perturbative QCD (pQCD) for not too large values of $n$. This type of sum-rules
has been, for a long time, the basis for precise determinations of the charm- and bottom-quark masses ($m_c$ and $m_b$)~\cite{Kuhn:2001dm,Kuhn:2007vp,Chetyrkin:2009fv,Chetyrkin:2010ic,Chetyrkin:2017lif,Dehnadi:2011gc,Dehnadi:2015fra}.

Recently, it has been shown that dimensionless ratios of roots of moments $M_{q,n}^{V}$ are also an
important source of reliable information about $\alpha_s$~\cite{Boito:2019pqp,Boito:2020lyp}. Given the
present status of the experimental measurements, and the fact that $\alpha_s$ extractions at low
energies often result in accurate predictions for $\alpha^{(n_f=5)}_s(m_Z)$, the ratios of charm-quark
moments lead to a particularly precise determination of the strong coupling. Ratios of this type had
already been exploited in determinations of $\alpha_s$ and $m_c$ from the pseudo-scalar current moments
by several lattice
groups~\cite{Petreczky:2019ozv,Allison:2008xk,McNeile:2010ji,Maezawa:2016vgv,Petreczky:2020tky}.

In all studies of this type, it is essential to reliably estimate the theoretical uncertainties associated with missing higher orders in the respective perturbative series. The pQCD expansion of the first three physical moments is known, at present, up to $\mathcal{O}(\alpha_s^3)$~\cite{Boughezal:2006px,Boughezal:2006uu,Chetyrkin:2006xg,Maier:2007yn,Maier:2008he,Sturm:2008eb,Maier:2009fz,Maier:2017ypu}.
The error stemming from lacking higher orders must therefore be carefully assessed through conservative renormalisation-scale variations
and/or estimates of higher-order coefficients.
Alternative treatments of these perturbative errors lead to discrepancies in the magnitude of the final uncertainties quoted by different groups~\cite{Dehnadi:2011gc,Dehnadi:2015fra,Chetyrkin:2009fv,Chetyrkin:2010ic,Chetyrkin:2017lif}.

In many cases, the final error on the extracted parameters receives an important contribution from the theoretical error associated with the truncation of perturbation theory. The appraisal of the different prescriptions for the computation of these errors can benefit from partial
knowledge about the yet unknown higher-order coefficients of the pQCD expansion of the moments $M_{q,n}^{V}$.
In this context, the large-$\beta_0$ limit of QCD is an important tool.
In this approximation, one first considers the limit of a large number of quark flavors, $n_f$, while keeping $\alpha_s n_f\sim \mathcal{O}(1)$.
The leading-$n_f$ terms of the pQCD series, which correspond to QED-like diagrams, are calculated to all orders in $\alpha_s$. Then, through the procedure known as naive
non-abelianization~\cite{Broadhurst:1994se,Beneke:1994qe,Beneke:1998ui}, the fermionic contribution to the leading-order (LO) QCD $\beta$ function
is replaced by the full coefficient, $\beta_0$, thereby
effectively introducing a set of non-abelian terms. This results in a series that is known to all orders in the coupling and whose Borel transform can be studied exactly.
The singularities of the Borel transform arising from IR and UV regions of loop subgraphs are the renormalons of perturbation theory, which govern the divergent
behaviour of the series at high orders. In QCD, IR renormalons play a particularly important role since in many cases they are in one-to-one correspondence with
non-perturbative QCD condensates arising in the operator product expansion.
In some situations, the large-$\beta_0$ limit provides a good estimate of higher-order coefficients. However,
even when this is not the case, it contains important information about the renormalons of perturbation theory, whose position is unchanged in the full QCD result.

The result for the small-momentum expansion of the vector correlator in the large-$\beta_0$ limit is available since the work of Grozin and Sturm~\cite{Grozin:2004ez}.
Here, we confirm their result and calculate, for the first time, the small-momentum expansion of the scalar, pseudo-scalar, and axial-vector correlators at
$\mathcal{O}(1/\beta_0)$. From a phenomenological point of view, the main focus is on the vector and pseudo-scalar correlators, since their small-
momentum expansion is the input for
the precise extraction of $m_c$, $m_b$, and $\alpha_s$ from data on the $R_{q\bar{q}}(s)$ ratio, in the vector case, and for the determination of $m_c$ and
$\alpha_s$ from lattice data for the
pseudo-scalar correlator. (Lattice data for the vector and axial-vector charm moments also exist, see e.g.~\cite{Allison:2008xk}, but are not as competitive as the lattice pseudo-scalar moments.)

Our results for the vector and pseudo-scalar correlators are then employed in a study of the
perturbative behaviour of the ratios of moments used for the extraction of $\alpha_s$. We obtain their
Borel transform in closed form, study their renormalon content, and show that these ratios benefit from
a partial cancelation of the leading UV renormalon, as well as a reduction of the leading IR pole residue.
This softening of the leading singularities is behind the good perturbative behavior of these moments.
Additionally, the knowledge of the renormalon singularities provides us with new information that can be used to design combinations of moments that exihibit stronger cancellations of the leading renormalons.

This work is structured as follows. In Sec.~\ref{sec:Theory} we define the correlators we are
interested in, their moments, and the ratios of moments. In Sec.~\ref{sec:MomentsM} we describe and present the calculation of the small-momentum expansion of the quark-current correlators at $\mathcal{O}(1/\beta_0)$. Then, in Sec.~\ref{sec:PtSeriesR}, these results are used to
obtain the \Lb expansion of the ratios of vector and pseudo-scalar moments employed in $\alpha_s$
analyses. We discuss the leading renormalon contribution to the ratios and show that partial cancellations take place, which is one of the main results of this paper. We also discuss
how to combine ratios of moments so as to obtain better-behaved perturbative series. Our conclusions are presented in Sec.~\ref{sec:Conclusions}. Finally, details about the
small-momentum expansion of the relevant two-loop integrals and a number of explicit results from our
calculations are relegated to Appendices~\ref{app:ExpLoop} and~\ref{app:ExplicitResults}, respectively.

\section{Theory overview}
\label{sec:Theory}
In this section we define the correlators that will be calculated in Sec.~\ref{sec:MomentsM} and discuss
their small-momentum expansion, which, in the vector case, is related to the moments of Eq.~\eqref{eq:MnV}.
We also define the dimensionless ratios of moments whose perturbative behaviour will be studied in Sec~\ref{sec:PtSeriesR}.

Even though our main focus is on the vector and pseudo-scalar correlators, given the phenomenological application of their small-momentum expansion as already discussed, for completeness we will present results for the vector ($V$), axial-vector ($A$), scalar ($S$), and pseudo-scalar ($P$) correlators
which we define as
\begin{equation}
(q^2 g_{\mu\nu}-q_\mu q_\nu )\Pi^\delta(s) - q_\mu q_\nu \Pi_{L}^\delta(s) = -i \!\int\! {\rm d}x\, e^{iq\cdot x} \sand{\Omega}{T j^\delta_\mu(x) j^{\delta\,\dagger}_\nu(0)}{\Omega}\,, \label{eq:VandACorrDef}
\end{equation}
for $\delta = V,A$ whereas
\begin{equation}
\Pi^\delta(s) = i \!\int \!{\rm d}x\, e^{iq\cdot x} \sand{\Omega}{T j^\delta(x) j^{\delta\,\dagger}(0)}{\Omega}\,,\label{eq:PandSCorrDef}
\end{equation}
for $\delta=S,P$. In the equations above $q^2=s$ and the bilinear quark currents are
\begin{align}
& j^V_\mu(x) = \bar{q}(x)\gamma_\mu q(x)\, ,& j^A_\mu(x)& = \bar{q}(x)\gamma_\mu \gamma_5 q(x)\,, \nn \\
& j^S(x) = 2m_q\,\bar{q}(x)q(x)\, , \quad {\rm and} &j^P(x)& = 2im_q\, \bar{q}(x)\gamma_5 q(x)\,. \label{eq:Currents}
\end{align}
The mass factor in the scalar and pseudo-scalar currents, which in this context corresponds to the bare mass, is introduced to ensure
renormalisation group invariance~\cite{Chetyrkin:1994js}.
The longitudinal contribution to the vector correlator $\Pi_L^V$ is zero due to the vector Ward identity. In the case of the axial-vector current,
$\Pi_L^A$ can be obtained by applying the projector $q^\mu q^\nu$ or using the axial Ward identity, which relates this contribution to the pseudo-
scalar correlator\cite{Maier:2009fz,Sturm:2008eb}.\footnote{The axial-vector moments are defined with respect to the small momentum expansion
of the transverse contribution.} When using dimensional regularisation for loop computations, the currents that contain $\gamma_5$ must be
carefully extended to $d$ dimensions; we employ the prescription described in Ref.~\cite{Larin:1993tp}.\footnote{No finite renormalisation of the
axial and pseudo-scalar currents is required in our case.}

With the usual definition of the experimentally accessible $R_{q\bar q}(s)$ ratio
\begin{equation}
\label{eq:Rqq}
R_{q\bar{q}}(s) = \frac{3s}{4\pi\alpha^2(s)}\sigma_{e^+e^-\to\, q\bar{q}\,+X}(s) \simeq
\dfrac{\sigma_{e^+e^-\to\, q\bar{q}\,+X}(s)}{\sigma_{e^+e^-\to\,\mu^+\mu^-}(s)}\,,
\end{equation}
where $\alpha$ is the effective electromagnetic coupling constant, the corresponding moments of Eq.~\eqref{eq:MnV} can be related to the coefficients of the Taylor expansion of the vector-current correlator around $s=0$ using analyticity and unitarity as
\begin{equation}
\label{eq:MqVTh}
M_{q,n}^{V} =\int_{s_{\rm th}}^{\infty} \frac{\df s}{s^{n+1}}R_{q\bar q}(s)=\dfrac{12\pi^2 Q_q^2}{n!}\,\dfrac{{\rm d}^n}{{\rm d}s^n}\Pi_q^V(s)\Bigr|_{s=0}\,.
\end{equation}
We will generalize this definition beyond the vector current and define the moments
\begin{equation}
\label{eq:MqdeltaTh}
M_{q, n}^{\delta} =\dfrac{12\pi^2 Q_q^2}{n!}\,\dfrac{{\rm d}^n}{{\rm d}s^n}\Pi_q^\delta(s)\Bigr|_{s=0}\,.
\end{equation}
As will be discussed in the next section, we restrict the analysis to physical moments, i.e.\ those
that do not require a scheme-dependent subtraction besides coupling and
mass renormalisation. For vector and axial-vector correlators this means $n\geq 1$. For $\delta=P,S$
in Eq.~\eqref{eq:MqdeltaTh} we must have $n\geq 0$.\footnote{Care must be taken when comparing with
other papers since in some cases the $2m_q$ factor is not included in the $S$ and $P$ quark currents
and a $q^2$ appears on the left-hand side of Eq.~\eqref{eq:PandSCorrDef}. Effectively, this shifts the
values of $n$ by one unit for $\delta=S$, $P$ and our moment $M_{q,n}^P$ corresponds to the moment with
the $n+1$ in the conventions of Ref.~\cite{Maier:2009fz}. (Here we follow more closely the definitions
of~\cite{Dehnadi:2011gc,Dehnadi:2015fra}.)} In all cases, the description in terms of standard
perturbative QCD supplemented with OPE condensate contributions breaks down for large values of $n$,
when a non-relativistic treatment becomes imperative, since the moments in this case are dominated by
the resonant contributions. Therefore, our phenomenological analysis will be restricted to values
of $n\leq 4$.

The expansion of the moments $M_{q,n}^\delta$ in perturbative QCD can be cast in the following general form
\begin{align}\label{eq:MqnPtExp}
M_{q, n}^{\delta} =\frac{1}{[2\overline m_q(\mu_m)]^{2n}} \sum_{i=0} \biggl[\frac{\alpha_s(\mu_\alpha)}{\pi}\biggr]^i
\sum_{a=0}^{i} \sum_{b=0}^{[i-1]} \, c^{\delta, (n)}_{i,a,b}\, \ln^a\!\biggl[\frac{\mu_m}{\overline m_q(\mu_m)}\biggr]
\ln^b\!\biggl[\frac{\mu_\alpha}{\overline m_q(\mu_m)} \biggr],
\end{align}
where we define $[i-1]\equiv {\rm max}(i-1,0)$, $\alpha_s(\mu_\alpha)\equiv \alpha_s^{(n_f)}(\mu_\alpha)$ and $\overline m_q(\mu_m)\equiv \overline m^{(n_f)}_q(\mu_m)$, with
$n_f$ the number of active quark flavours.\footnote{In full QCD one has $n_f=n_\ell + 1$, with $n_\ell$ the number of massless quarks, but since heavy-quark mass loops are $1/\beta_0$ suppressed, in the large-$\beta_0$ one effectively has $n_f=n_\ell$.} Here $\overline m_q(\mu_m)$ and $\alpha_s(\mu_\alpha)$ are
the quark mass and strong coupling, respectively, in the $\MSb$ scheme. The independent (or non-log) coefficients
$c^{\delta, (n)}_{i,0,0}$ must be calculated in perturbative QCD, while the logarithms can be
generated with renormalisation group equations. For notational simplicity we
also omit the quark charge dependence (through a global factor of $9Q_q^2/4$) and
the $n_f$ dependence of the coefficients $c^{\delta, (n)}_{i,0,0}$. The
expansion is exactly known in QCD up to $\mathcal{O}(\alpha_s^3)$ for the first three physical moments for the four
correlators we consider here thanks to a huge computational
effort~\cite{Boughezal:2006px,Boughezal:2006uu,Chetyrkin:2006xg,Maier:2007yn,Maier:2008he,Sturm:2008eb,Maier:2009fz}. The fourth moment of the pseudo-scalar and vector currents are also known exactly~\cite{Maier:2009fz, Maier:2017ypu} while higher moments have been estimated~\cite{Hoang:2008qy, Kiyo:2009gb, Greynat:2010kx, Greynat:2011zp}.
To be fully general, we
allow for different renormalisation scales in the mass and the coupling. The leading
logarithm in Eq.~\eqref{eq:MqnPtExp} appears already at order~$\alpha_s$.

Only the vector moments can be determined from experimental data. Sum rules with the vector moments of Eq.~\eqref{eq:MqVTh} are the basis for precise extractions of
$m_c$ and $m_b$ from $R_{q\bar{q}}(s)$ experimental
data~\cite{Kuhn:2001dm,Kuhn:2007vp,Chetyrkin:2009fv,Chetyrkin:2010ic,Chetyrkin:2017lif,Dehnadi:2011gc,Dehnadi:2015fra}. The first few charm pseudo-scalar moments have been determined from lattice simulations with good precision by several
groups~\cite{Allison:2008xk,McNeile:2010ji,Maezawa:2016vgv,Petreczky:2019ozv,Petreczky:2020tky} and analogous sum rules for the pseudo-scalar moments have been used
in the extraction of $m_c$ from these lattice results. The 0-th pseudo-scalar moment, due to its reduced mass dependence, has also been used for $\alpha_s$ determinations.

It is also useful to work with dimensionless ratios of roots of moments (with $n>0$). In these ratios
the mass dependence almost completely disappears, entering only through $\alpha_s^2$-suppressed logarithms.
We define the following dimensionless ratios
\begin{equation}\label{eq:RqnDef}
R_{q,n}^{\delta}\equiv \frac{\bigl(M_{q,n}^{\delta}\bigr)^\frac{1}{n}}{\bigl(M_{q,n+1}^{\delta}\bigr)^\frac{1}{n+1}}\,,
\end{equation}
where $\delta=V,P$. This type of ratios of moments was first introduced for the analysis of pseudo-scalar lattice data~\cite{Maezawa:2016vgv,Petreczky:2019ozv}.
Their use in the case of the vector current was introduced in Refs.~\cite{Boito:2019pqp,Boito:2020lyp} where it was shown that they can be employed
for precise extractions of $\alpha_s$ thanks to their reduced mass dependence and to the fact that these ratios can be accurately determined from $R_{q\bar q}(s)$ experimental data, benefiting from positive correlations between the moments $M_{q,n}^{V}$ and $M_{q,n+1}^{V}$.

The general structure of the perturbative expansion of the moments $R_{q,n}^{\delta}$ is
\begin{align}\label{eq:RqnPTExp}
R^{\delta}_{q,n} =
\sum_{i=0} \bigg[\frac{\alpha_s(\mu_\alpha)}{\pi}\bigg]^i \sum_{k=0}^{[i-1]}\sum_{j=0}^{[i-2]} r^{\delta, (n)}_{i,j,k}
\ln^j\!\biggl[\frac{\mu_m}{\overline m_q(\mu_m)}\biggr]\!
\ln^k\!\biggl[\frac{\mu_\alpha}{\overline m_q(\mu_m)}\biggr],
\end{align}
where the mass dependence in the prefactor of $M_{q,n}^\delta$ is explicitly canceled by construction and the coefficients $r^{\delta,(n)}_{i,j,k}$ can be obtained from the
$c^{\delta, (n)}_{i,j,k}$ upon re-expansion of the ratios. Since the ratios are dimensionless, the residual mass dependence appears only in the arguments of the logarithms, and
now start to contribute only at $\mathcal{O}(\alpha_s^2)$~\cite{Boito:2019pqp}.
When comparing the results in \Lb and QCD it will be convenient to consider the scale dependent $\alpha_s$ coefficients of Eq.~(\ref{eq:RqnPTExp}) that we define as
\begin{equation}\label{eq:bar_rCoeff}
\bar r_{i,n}^\delta (\mu_\alpha,\mu_m)=\sum_{k=0}^{[i-1]}\sum_{j=0}^{[i-2]} r^{\delta, (n)}_{i,j,k}
\ln^j\!\biggl[\frac{\mu_m}{\overline m_q(\mu_m)}\biggr]\!
\ln^k\!\biggl[\frac{\mu_\alpha}{\overline m_q(\mu_m)}\biggr].
\end{equation}

Finally, we remark that the dimensionless combinations of moments are certainly not unique. In fact, with the knowledge about the renormalon
singularities in \Lb obtained here, we are in a position to design other dimensionless combinations of moments that could display a better
perturbative behaviour due to stronger renormalon cancellation. We discuss this possibility in Sec.~\ref{sec:NewRatios}.

\section{The moments \boldmath \texorpdfstring{$M_{q,n}^{\delta}$}{Mndelta} in the large-\texorpdfstring{$\beta_0$}{beta0} limit }
\label{sec:MomentsM}
In this section we will present the results for the small-momentum expansion of the vector, axial-vector, scalar, and pseudo-scalar correlators in the large-$\beta_0$ limit of QCD. We will cast the expansion of the renormalised correlators in this limit in the following form
\begin{equation}
\widehat \Pi^\delta(q^2) = \frac{N_c}{16\pi^2} \sum_{n=n_\delta}^\infty \bigg[\frac{s}{4\overline m_q^2(\mu)} \biggr]^n N_n^\delta\, C_n^\delta(\mu)\,,\label{eq:Piexpansion}
\end{equation}
where $N_c=3$ is the number of colours and $N_n^\delta$ is the $\mathcal{O}(\alpha_s^0)$ (one-loop) result in $d=4$
dimensions.\footnote{Specifically, with our conventions we have, for the vector case, $N_1^V=16/15$, $N_2^V=16/35$, and $N_3^V=256/945$. (With the conventions of
Ref.~\cite{Grozin:2004ez} the $N_n^V$ would be divided by $4^n$.) For the pseudo-scalar moments we have $N_0^P=4/3$,
$N_1^P=8/15$, and $N_2^P=32/105$. The one-loop normalization for $S$ and $A$ moments can be found in the accompanying file~\cite{gitlabFile}.} With
this normalisation, the perturbative expansion of $C_n^\delta(\mu)$ starts as $1$. We are interested in physical moments, i.e.\ those that
do not have an UV divergence after coupling and mass renormalisation which would require a scheme-dependent subtraction. Accordingly,
we remove from the definition of $\widehat \Pi^\delta(q^2)$ in Eq.~\eqref{eq:Piexpansion} the unphysical terms setting $n_A=n_V=1$ and $n_S=n_P=0$. The moments are characterised by
the non-trivial reduced moments $C_n^\delta(\mu)$, for which we will obtain a Borel representation. They retain a quark-mass dependence through the ratio $\mu/\mbar_q(\mu)$, which appears in logarithms in the perturbative expansion.
From the definition of the moments $M_{q,n}^{\delta}$
given in Eq.~\eqref{eq:MqVTh}, one obtains
\begin{equation}\label{eq:MqndeltaCndelta}
M_{q,n}^{\delta} = \frac{9}{4}Q_q^2\, \frac{N_n^\delta}{[4\overline m_q^2(\mu)]^n} C_n^\delta(\mu)\,.
\end{equation}
For the calculation of $C_n^\delta(\mu)$ in the large-$\beta_0$ limit, given that renormalisation is required, we rely on the formalism described in detail in Ref.~\cite{Grozin:2003gf}, which was employed in the original calculation of the small-momentum expansion of the vector correlator in this limit~\cite{Grozin:2004ez}. (This formalism was recently generalised to the case of quantities with cusp anomalous dimension in~\cite{Gracia:2021nut}.)

\begin{figure}[!t]
\begin{center}
\includegraphics[width=0.45\textwidth]{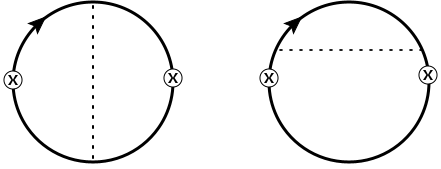}
\caption{Feynman diagrams for the calculation of the heavy-quark correlators in the large-$\beta_0$ limit. The rightmost diagram must be counted twice. Dashed lines represent gluon propagators with light-quark bubble insertions. Crosses stand for the insertion of the currents of Eq.~\eqref{eq:Currents}. }
\label{fig:two-loop}
\end{center}
\end{figure}

To obtain $C_n^\delta(\mu)$ in the large-$\beta_0$ limit one starts from the insertion of massless quark bubbles in the gluon propagators that appear in two-loop diagrams, as depicted in Fig.~\ref{fig:two-loop}. The insertion of these fermion loops amounts, essentially, to the calculation of the two-loop correction with the gluon propagator in the Landau gauge analytically regularized~\cite{Beneke:1998ui,Grozin:2003gf}.

Quite generally, a Borel representation for the renormalised functions $C_n^\delta(\mu)$ can be written in the following form
\begin{equation}
C_n^\delta(\mu) = 1 + \frac{1}{\beta_0}\Biggl[\int_0^{\alpha_s(\mu)}\! \frac{{\rm d}\alpha}{\alpha} \biggl( \frac{2\pi\gamma(\alpha)}{\alpha} - \frac{\gamma_0}{2} \biggr)
+\int_0^\infty \!{\rm d}u \, e^{-\frac{u}{a_\mu}} S_n^\delta(u)\!\Biggr] \!+ \mathcal{O}\!\biggl(\frac{1}{\beta_0^2}\biggr),\label{eq:Cndelta}
\end{equation}
where
\begin{equation}
a_\mu=\frac{\beta_0 \alpha_s(\mu)}{4\pi}\,,\label{eq:beta}
\end{equation}
with $\beta_0$ the one-loop coefficient in the perturbative expansion of the QCD $\beta$ function, defined as
\begin{equation}
\mu \frac{\df \alpha_s (\mu)}{\df \mu} = - 2 \alpha_s (\mu) \sum_{n = 0} \beta_n\biggl[ \frac{\alpha_s (\mu)}{4 \pi} \biggr]^{n+1} \!\equiv
\beta(\alpha_s (\mu))\,.
\end{equation}
In the conventions we are following $\beta_0 = 11\,N_c/3 -4T_F n_\ell/3$, where $T_F=1/2$ and $n_\ell$ is the number of light-quark flavors. We remind that the running of $\alpha_s(\mu)$ is to be performed with one-loop accuracy.

The first integral in Eq.~\eqref{eq:Cndelta} over $\gamma$, the anomalous dimension of $C_n^\delta(\mu)$, is present only in quantities that require additional subtractions beyond the massless fermion bubble renormalisation in the dressed gluon propagator~\cite{Grozin:2003gf,Beneke:1998ui}.
Here, besides the coupling renormalisation, the renormalisation $\MSb$-mass factor in the expansion brings an extra renormalisation constant,\footnote{Since we express the bare quark mass in terms of the $\MSb$ mass, in practice this amounts to dropping all $1/\varepsilon^n$ divergent terms in the series. The dropped factor is precisely $Z_m^{-2n}-1\approx -2n(Z_m - 1)$.}
given by $Z_m^{2n}$, and therefore the anomalous dimension for the quantities $C_n^\delta(\mu)$ is
\begin{equation}\label{eq:oldBeta}
\gamma(\alpha) =-4n\,\gamma_m(\alpha)\,,
\end{equation}
where $\gamma_m(\alpha)$ is the $\MSb$ mass anomalous dimension at $\mathcal{O}(1/\beta_0)$ accuracy~\cite{PalanquesMestre:1983zy, Grozin:2003gf}
\begin{equation} \label{Eq. gamma_m}
\gamma_m(\alpha) = -\frac{C_{\!F} a_\mu (3 + 2 a_\mu) \Gamma (4 + 2 a_\mu)}{ 3\beta_0(2 +a_\mu) \Gamma (1 - a_\mu) \Gamma (2 + a_\mu)^3 }
+ \mathcal{O}\!\biggl(\frac{1}{\beta_0^2}\biggr),
\end{equation}
and $a_\mu$ is given in Eq.~\eqref{eq:beta}. Our definition of the mass anomalous dimension is
\begin{equation}
\frac{\mu}{\overline{m}_q (\mu)} \frac{\df \overline{m}_q(\mu)}{\df \mu} = 2 \gamma_m[\alpha_s(\mu)]=2\!\sum_{k=0}\gamma^{(k)}_m\biggl[\frac{\alpha_s(\mu)}{4\pi}\biggr]^{\!k},
\end{equation}
with $\gamma^{(0)}_m = - 3 C_F=-4$. In Ref.~\cite{Gracia:2021nut} a recursive formula to efficiently obtain $\gamma^{(k)}_m$ was provided.
The solution to the RG equation in the large-$\beta_0$ limit is simple and if expanded strictly to $\mathcal{O}(1/\beta_0)$ can be written as
\begin{equation}
\overline{m}_q(\mu) = \overline{m}_q \biggl\{ 1 - \frac{1}{\beta_0} \int_{\alpha_s(\overline m_q)}^{\alpha_s(\mu)}
\frac{\df \alpha}{\alpha} \biggl( \frac{4 \pi \gamma_m (\alpha)}{\alpha} -
\gamma_m^{(0)} \biggr)\! - \frac{\gamma_m^{(0)}}{\beta_0} \log \biggl[
\frac{\alpha_s(\mu)}{\alpha_s(\overline m_q)} \biggr] \!\biggr\},\label{eq:mqrunning}
\end{equation}
where here and in what follows $\mbar_q \equiv \mbar_q(\mbar_q)$.

The functions $S_n^\delta(u)$ in Eq.~\eqref{eq:Cndelta} are, therefore, the Borel transforms of $C_n^\delta(\mu)$.
From the Borel representation of the functions $C_n^\delta(\mu)$ it is straightforward to extract their $\alpha_s$ expansion in the large-$\beta_0$ limit as
\begin{equation}\label{eq:CndeltaPtCoeff}
C_n^\delta(\mu) = \!\Biggl[1 + \frac{1}{\beta_0}\sum_{k=1}^\infty\Biggl(\frac{{\rm d}^{k-1}S_n^\delta}{{\rm d}u^{k-1}}
\bigg|_{u=0}\beta_0^k-\frac{2n\gamma^{(k)}_m}{k}\Biggr)\!\biggl[\frac{\alpha_s(\mu)}{4\pi} \biggr]^{\!k} + \mathcal{O}\!\biggl(\frac{1}{\beta_0^2} \biggr)\!\Biggr].
\end{equation}

Explicit analytic expressions for $S_n^\delta(u)$ are obtained as
\begin{equation}
S_n^\delta(u) = \frac{F_n^\delta(0,u) - F_n^\delta(0,0)}{u}\,,
\end{equation}
where the auxiliary functions $F_n^\delta(\varepsilon,u)$ are given by~\cite{Grozin:2003gf}
\begin{equation}
F_n^\delta(\varepsilon,u) = u\, e^{\gamma_E \varepsilon} a_n^{\delta}(1+u-\varepsilon,\varepsilon)\mu^{2u} D(\varepsilon)^{\frac{u}{\varepsilon} -1}.
\end{equation}
In the last expression, $D(\varepsilon)$ is the massless fermionic correction to the gluon propagator in $d=4-2\varepsilon$ dimensions and
$a_n^{\delta}(1+u-\varepsilon, \varepsilon)$ are the coefficients in the small-momentum expansion of the two-loop correction with the Landau-gauge gluon propagator analytically regularised, i.e.\ with the denominator $1/(-p^2)$ modified to $1/(-p^2)^{(1+u-\varepsilon)}$
\begin{equation}
a^{\delta}(1+u-\varepsilon,\varepsilon) = \sum_n \biggl(\frac{q^2}{4 m_q^2} \biggr)^{\!\!n} \,N_n^\delta(\varepsilon)\, a_n^{\delta}(1+u-\varepsilon, \varepsilon)\,, \label{eq:a1n}
\end{equation}
where $N_n^\delta(\varepsilon)$ ensures the result of $a_n^\delta$ is normalised to the LO result and at this point $m_q$ is still the bare mass.

The result of Eq.~\eqref{eq:a1n} is obtained computing the Feynman diagrams shown in Fig.~\ref{fig:two-loop}.
After calculating the Dirac trace, all terms in the numerator can be written in terms of propagators, which reduces
the problem to the study of scalar two-loop integrals given explicitly in Eq.~\eqref{eq:J2} of Appendix~\ref{app:ExpLoop}.
The scalar two-loop integrals are then expanded around $q^2=0$ using the method of Ref.~\cite{Davydychev:1992mt} as described in detail in Appendix~\ref{app:ExpLoop} and, after setting $q^2=0$, one is left with single-scale tadpole integrals that can be solved analytically.

\subsection{Results}
Following the procedure outlined above, we performed the calculation of the small-momentum expansions of vector, axial-vector, scalar, and pseudo-scalar correlators in the large-$\beta_0$ limit.

For the vector correlator, we computed the functions $S_n^V(u)$ up to $n=12$, finding agreement with the results presented in Ref.~\cite{Grozin:2004ez}, which were given up to $n=2$. The results for the scalar, pseudo-scalar, and axial-vector current correlators are obtained here for the first time.
Here we quote explicitly the results for the first three physical moments of each current, but obtaining the functions $S_n^\delta(u)$ for higher values of $n$ is, essentially, just a matter of computational time. (We remind that for the vector and axial-vector current correlators $n$ starts at 1, while for the pseudo-scalar and scalar correlators $n$ starts at 0.)

The results can be conveniently cast in terms of polynomials of $u$, $P_n^\delta(u)$, which must be determined case by case, in the following form
\begin{subequations}
\label{eq:Sn}
\begin{align}
& S_n^V(u) = \frac{6C_Fn}{u}
-3C_F\biggl[\frac{e^{5/3}\mu^2}{\overline m_q^2(\mu)} \biggr]^u\frac{ 4^n \Gamma (2-u) \Gamma (u) \Gamma (2+n+u)^2}{(n+u) \Gamma (3+2 n+2 u)}P_n^V(u)\,, \label{eq:SnV}\\
& S_n^A(u) = \frac{6C_Fn}{u}
-3C_F\biggl[\frac{e^{5/3}\mu^{2}}{\overline m_q^2(\mu)} \biggr]^u\frac{4^{n} \Gamma (2-u) \Gamma (u) \Gamma (2+n+u)^2}{(n+u) (1+n+u) \Gamma (3+2 n+2 u)}P_n^A(u)\,,\label{eq:SnA} \\
& S_n^S(u) = \frac{6C_Fn}{u}
-3C_F\biggl[\frac{e^{5/3}\mu^{2}}{\overline m_q^2(\mu)} \biggr]^u\frac{4^{n} \Gamma (2-u) \Gamma (u) \Gamma (1+n+u)^2}{(3+2 n+2 u) \Gamma (2+2 n+2 u)}P_n^S(u)\,,\label{eq:SnS}\\
& S_n^P(u) = \frac{6C_Fn}{u}
- 3C_F\biggl[\frac{e^{5/3}\mu^{2}}{\overline m_q^2(\mu)} \biggr]^u \frac{4^{n} \Gamma (2-u) \Gamma (u) \Gamma (2+n+u)^2}{(1+n+u) \Gamma (3+2 n+2 u)} P_n^P(u)\,.\label{eq:SnP}
\end{align}
\end{subequations}
The first few polynomials $P_n^\delta(u)$ are available in Appendix~\ref{app:Polynomials}. Additional results can be found in the accompanying file~\cite{gitlabFile}. As we are working at leading order in
$1/\beta_0$, one can replace $\overline m_q(\mu)$ by $\overline m_q$ in these relations, since the running of the quark mass produces terms that are $1/\beta_0^2$ and beyond, as per Eq.~(\ref{eq:mqrunning}). With this replacement it is easy to show exact $\mu$-independence of the Borel integrals of the moments in the
large-$\beta_0$ limit.

The general structure of the functions $S_n^\delta(u)$ fulfils the expectations of typical results in large-$\beta_0$. Terms with the factor
$[e^{5/3}\mu^2/\overline m_q^2(\mu)]^u$ lead to a Borel integral that is scheme and scale invariant~\cite{Beneke:1998ui}. However, here, since
renormalisation is required, the functions $S_n^\delta(u)$ have a $1/u$ term without this factor, which is a reminder of the renormalisation
scheme and scale dependence of the quark mass\cite{Beneke:1994qe,Beneke:1998ui}. In fact, quite generally, this first term can be written as
$-2n\,\gamma_m^{(0)}/u$.
There is, however, no singularity at $u=0$ thanks to an exact cancellation when both terms in Eqs.~\eqref{eq:Sn} are added up. The scheme and scale dependence arising from the $1/u$ term is canceled by the integral over the anomalous
dimension and the global mass prefactor in Eq.~\eqref{eq:MqndeltaCndelta}.

There are several non-trivial tests that we have performed to ensure the correctness of our results:
\begin{itemize}
\item In all cases, the leading-$n_\ell$ power at each order in $\alpha_s$ in the perturbative expansion of the moments $M_{q,n}^\delta$ should be correctly reproduced. We have checked that this is the case for results that are known in QCD from Refs.~\cite{Chetyrkin:2006xg,Maier:2007yn,Maier:2008he,Sturm:2008eb,Maier:2009fz}.
\item The functions $S_n^\delta(u)$ written in terms of the $\MSb$ quark mass have simple poles of IR origin on the positive $u$ axis at $u=2,3,4,\ldots$; no pole at $u=1$ is present. This is expected, since the leading condensate contribution is the dimension-$4$ gluon condensate~\cite{Beneke:1998ui}, which corresponds to the pole at $u=2$. (We have checked that rewriting these Borel transforms in the on-shell scheme, the pole mass renormalon at $u=1/2$ becomes the leading IR singularity, followed by an additional $u=1$ pole, again as expected~\cite{Beneke:1994qe}. The contribution of the pole mass to the IR singularities is however not related to the OPE condensates.)
\item A third rather non-trivial test is also related to the gluon-condensate contribution. The gluon condensate coefficient is known for the four
currents at NLO~\cite{Broadhurst:1994qj}. In one specific case, namely the moment $n=2$ of the pseudo-scalar correlator, this coefficient vanishes
at lowest order. Accordingly, we find that for the $S_2^{P}(u)$, and only in this case, the IR singularity at $u=2$ is absent, because
$P_2^P(u)$ has a zero at $u=2$ as can be seen in Eq.~\eqref{eq:PP}, in agreement with the expectation that the IR renormalons in the $\MSb$
scheme are in one-to-one correspondence with OPE contributions.
\item Finally, we have verified that the axial Ward identity relating the longitudinal part of $\Pi_L^A(u)$ and $\Pi^P(u)$ is verified at $\mathcal{O}(1/\beta_0)$.
\end{itemize}

Because of the existence of the IR poles, the Borel integral in Eq.~\eqref{eq:Cndelta} is not well
defined and a prescription to deal with the singularities along the positive real axis must be
adopted. Here we use the principal value prescription, such that the Borel integral
acquires an imaginary part whose value (divided by $\pi$) is commonly considered to be a good
estimate for the ambiguity of the Borel integral.
The contribution of each pole to this
ambiguity scales as a non-perturbative correction. At the scales we consider here, the ambiguity of the Borel integral is
numerically quite small, as we will show in the next sections, which simply reflects the fact that the
non-perturbative corrections in the OPE, dominated by the gluon condensate contribution, are rather
small. This is particularly true for bottom quark moments, where the non-perturbative contributions can
be neglected for all practical purposes~\cite{Dehnadi:2015fra,Boito:2020lyp}.

Apart from the IR renormalon poles that we already mentioned, the functions $S_n^\delta(u)$ have UV
poles at $u=-1,-2,-3,\ldots$ as well, which lie on the negative real axis. In the functions
$S_n^\delta(u)$ all IR singularities are simple poles, stemming from $\Gamma(2-u)$.
For the UV poles the pattern that emerges is a little more intricate.
There are poles at negative $u$
from $\Gamma(u)$ as well as from squared gamma functions in the numerator.
The UV poles can be simple or double plus simple depending on the structure of the denominator.
The functions $S_n^V(u)$, for instance, have singularities with a double- plus simple-pole structure at $u=-n,-
(n+2),-(n+3),\ldots$ while all other UV poles are simple. For $\delta = P,S$ the double poles start at $u = - (n+1)$. There can be
exceptions, though. For example, $u=-7$ is a root of $P_0^P(u)$ [\,see Eq.~\eqref{eq:PP}\,], and the UV pole at $u=-7$ becomes
simple in $S_0^P(u)$. We will not speculate about the physical origin of
this pattern, but the leading UV renormalons will be discussed in more
detail in the context of the ratios $R_{q,n}^\delta$ in the remainder of this paper.

Our calculation of the moments $M_{q,n}^\delta$ for $\delta=P$, $S$, $A$ in the \Lb limit is a new result in the literature. From the expansion of these
results one can obtain their perturbative expressions in \Lb and read off the coefficients of the $\alpha_s^k\,n_\ell^{k-1}$ terms, which must
be the same as in full QCD. To expand the various gamma functions efficiently one can use the following compact form, valid for $n\geq0$
\begin{equation}
\Gamma(n+u)= (n-1)! \exp\biggl\{\!u \Bigl(H^{(1)}_{n-1}-\gamma_E \Bigr) + \sum_{k=2}^\infty \biggl[\frac{(-u)^k }{k}\Bigl(\zeta_k-H_{n-1}^{(k)}\Bigr)\!\biggr]\!\biggr\}\,,
\end{equation}
with $H_n^{(k)}\equiv \sum_{i=1}^n n^{-k}$ the harmonic number of order $k$ and $\gamma_E$ Euler's constant. Using the formula above, all gamma functions
appearing in Eq.~\eqref{eq:Sn} and the
$\mu$-dependent prefactor can be combined into a single exponential, which is afterwards expanded using Eq.~(6.5) of Ref.~\cite{Gracia:2021nut}. Finally, the expanded
exponential is easily combined with the (already expanded) accompanying finite polynomials into a single expansion using
\begin{equation}
\sum_{i = n} a_i x^i \sum_{j = m}^N b_j x^j =\sum_{i = n + m} x^i \sum_{j = m}^{\min (N, i - n)} a_{i - j} b_j \,,
\end{equation}
where both sums over $i$ run all the way to infinity. Exemplarily we work out analytically the main steps of the expansion for $S_n^V(u)$:
\begin{align}
\frac{S_n^V(u)}{3C_F}= & \frac{2n}{u} - \frac{4^{n} [(n+1)!]^2}{n u (2 n+2)!} \exp\biggl\{\!u \biggl[2 (H_{n+1}^{(1)}-H_{2 n+2}^{(1)})+\frac{2}{3}-\frac{1}{n}
+\log\biggl(\frac{\mu^2}{\overline m_q^2(\mu)}\biggr)\!\biggr] \\
& + \sum_{k=2}^\infty \frac{u^k}{k}\biggl[\frac{(-1)^k}{n^k}-1-2 (-1)^k H_{n+1}^{(k)}+(-2)^k H_{2 n+2}^{(k)} \nonumber\\
& +\Bigl((-1)^k \bigl(3-2^k\bigr)+1\Bigr) \zeta_k\biggr] \biggr\} P_n^V(u) \,,\nn
\end{align}
which implies the constraint $P_n^V(0)=2^{1-2 n} n^2 (2 n+2)!/[(n+1)!]^2$ satisfied by Eq.~\eqref{eq:Pvec}. The asymptotic expansion concerning IR poles is
\begin{align}\label{eq:asy}
\frac{S_n^V(u) \big|_{\rm IR}}{3C_F}\asymp\,& 4^n \!\sum_{m=2}^\infty \biggl[-\frac{e^{5/3}\mu^2}{\overline m_q^2(\mu)} \biggr]^m\frac{(m-1) [(1+m+n)!]^2 }{(m+n) (2+2 m+2 n)!}
\frac{P_n^V(m)}{u-m}\,, \\
\frac{S_n^A(u) \big|_{\rm IR}}{3C_F}\asymp\,&4^n\! \sum_{m=2}^\infty \biggl[-\frac{e^{5/3}\mu^2}{\overline m_q^2(\mu)} \biggr]^m\frac{ (m-1) [(m+n)!]^2}{2(m+n)(1+2 m+2 n)!}
\frac{P_n^A(m)}{u-m}\,, \nonumber\\
\frac{S_n^S(u) \big|_{\rm IR}}{3C_F}\asymp\,&4^n \! \sum_{m=2}^\infty \biggl[-\frac{e^{5/3}\mu^2}{\overline m_q^2(\mu)} \biggr]^m\frac{ (m-1) [(m+n)!]^2 }{(3+2m+2n) (1+2 m+2 n)!}
\frac{P_n^S(m)}{u-m}\,,\nonumber\\
\frac{S_n^P(u) \big|_{\rm IR}}{3C_F}\asymp\,&4^n\! \sum_{m=2}^\infty \biggl[-\frac{e^{5/3}\mu^2}{\overline m_q^2(\mu)} \biggr]^m\frac{(m-1) [(m+n)!]^2 }{2 (1+2 m+2 n)!}
\frac{P_n^P(m)}{u-m}\,,\nn
\end{align}
where $\asymp$ means singular part of. The above expressions are very useful to carry out the Borel integral with the principal value prescription, and are responsible for the total ambiguity of the Borel sum. Due to the more complicated pattern of UV singularities it is not easy to find the corresponding asymptotic expansions for arbitrary values of $n$.

We collect in Appendix~\ref{app:leading_nl} the results for the leading $n_\ell$ terms in the $\alpha_s$
expansions of the combinations $N_n^\delta C_n^\delta$ of Eq.~\eqref{eq:MqndeltaCndelta} up to $\mathcal{O}(\alpha_s^4)$, which is the first unknown
previous to this work in the case of $\delta=P$, $S$, and $A$. Additional terms in these expressions can be easily generated from the results
presented here.

\subsection[The ratios \texorpdfstring{$R_{q,n}^\delta$}{Rndelta} in the large-\texorpdfstring{$\beta_0$}{beta0} limit]{The ratios \boldmath \texorpdfstring{$R_{q,n}^\delta$}{Rndelta} in the large-\texorpdfstring{$\beta_0$}{beta0} limit}\label{eq:ratiosLarge}

We turn now to a discussion of the moment ratios $R_{q,n}^\delta$ in the \Lb limit. Using the definition of the dimensionless ratios, Eq.~\eqref{eq:RqnDef}, together with the Borel representation of $C_n^\delta(\mu)$ in Eq.~\eqref{eq:Cndelta}, by consistently re-expanding in $1/\beta_0$ one can obtain the Borel representation
of $R_{q,n}^\delta$ in the \Lb limit as
\begin{equation} \label{eq:RnFullResult}
R_{q,n}^\delta = \biggl(\frac{9}{4}Q_q^2 \biggr)^{\!\frac{1}{n(n+1)}}\frac{(N_n^\delta)^\frac{1}{n}}{(N_{n+1}^\delta)^\frac{1}{n+1}}
\biggl[ 1 + \frac{1}{\beta_0}\int_0^\infty\! {\rm d}u \, e^{-\frac{u}{a_\mu}} B_n^\delta(u) + \mathcal{O}\biggl(\frac{1}{\beta_0^2}\biggr)\! \biggr],
\end{equation}
where $B_n^\delta(u)$ are the Borel transforms of $R_{q,n}^\delta$, which can easily be written
in terms of the
$S_n^\delta(u)$ functions as follows
\begin{equation}\label{eq:Bndelta}
B_n^\delta(u) = \frac{S_n^\delta(u)}{n} - \frac{S_{n+1}^\delta(u)}{n+1}\,.
\end{equation}
In obtaining the above result we are tacitly assuming that the non-perturbative corrections in the OPE are smaller than the perturbative contribution such that they can be expanded out.
The knowledge of the gluon-condensate contribution shows that this is an excellent approximation for the bottom and a very good one for the charm~\cite{Boito:2019pqp,Boito:2020lyp}

Since the ratios $R_{q,n}^\delta$ are designed so as to cancel the explicit mass factor of Eq.~\eqref{eq:MqnPtExp}, their Borel transforms do not have
the term proportional to $\gamma^{(0)}_m/u$ which vanishes in Eq.~\eqref{eq:Bndelta}. Accordingly, the integral over the mass anomalous
dimension in Eq.~\eqref{eq:Cndelta} also vanishes, and the integral of ratios of moments are scheme and scale invariant thanks to the now global factor
of $[e^{5/3}\mu^2/\overline m_q^2(\mu)]^u$.
In the perturbative expansion, the residual mass dependence starting at $\mathcal{O}(n_\ell\alpha_s^2)$
now enters only through $1/\beta_0$-suppressed logarithms.
An important comment is that changing the renormalisation scale (or scheme) of the running quark mass brings corrections of order $1/\beta_0^2$ and superior, which
are subleading in our approximation and should consistently be dropped in a strict \Lb expansion.

\begin{figure}[!t]
\begin{center}
\subfigure[]{\includegraphics[width=0.31\textwidth]{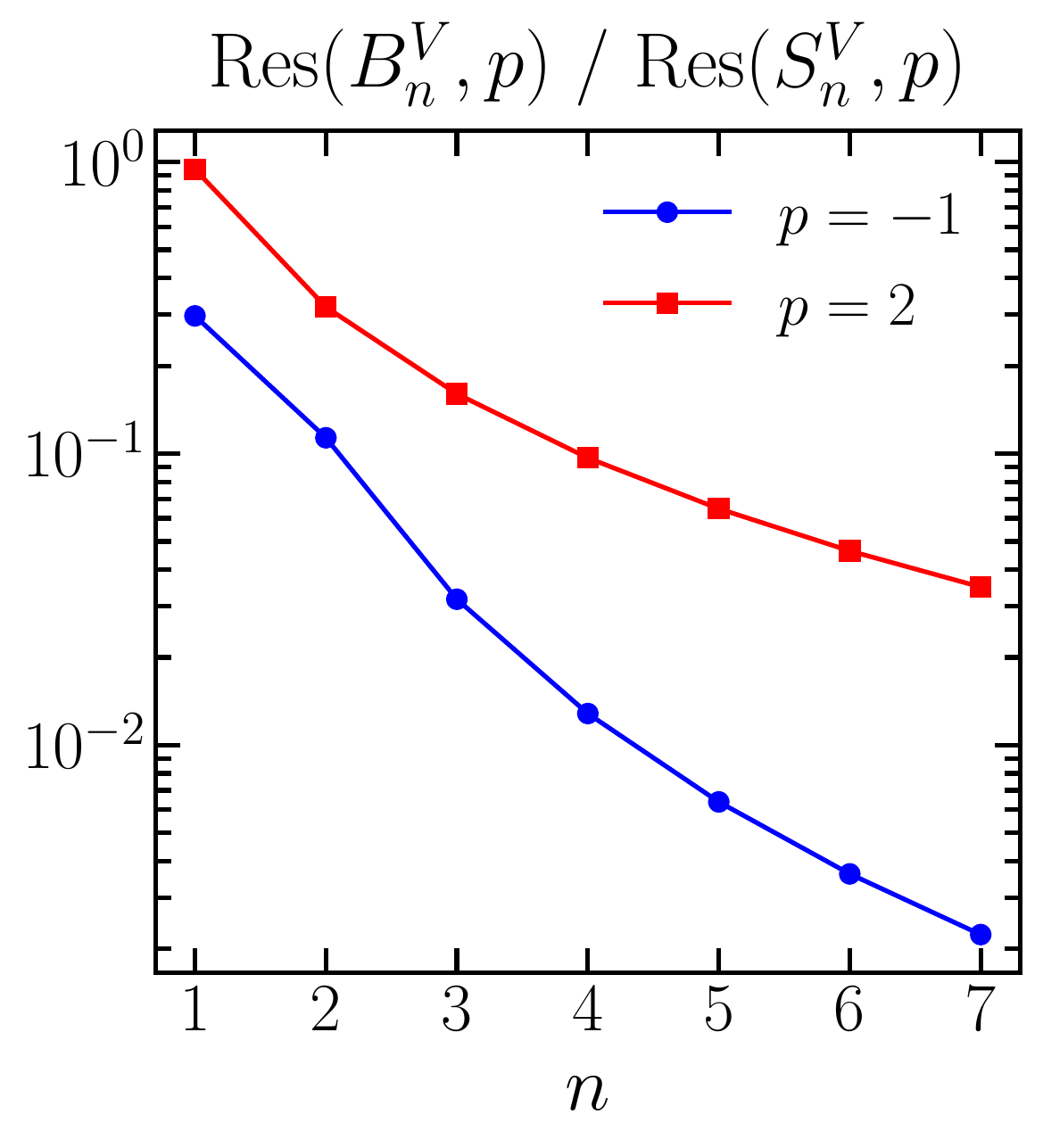}{\label{fig:Residues1}}}~~~~~~~~~
\subfigure[]{\includegraphics[width=0.31\textwidth]{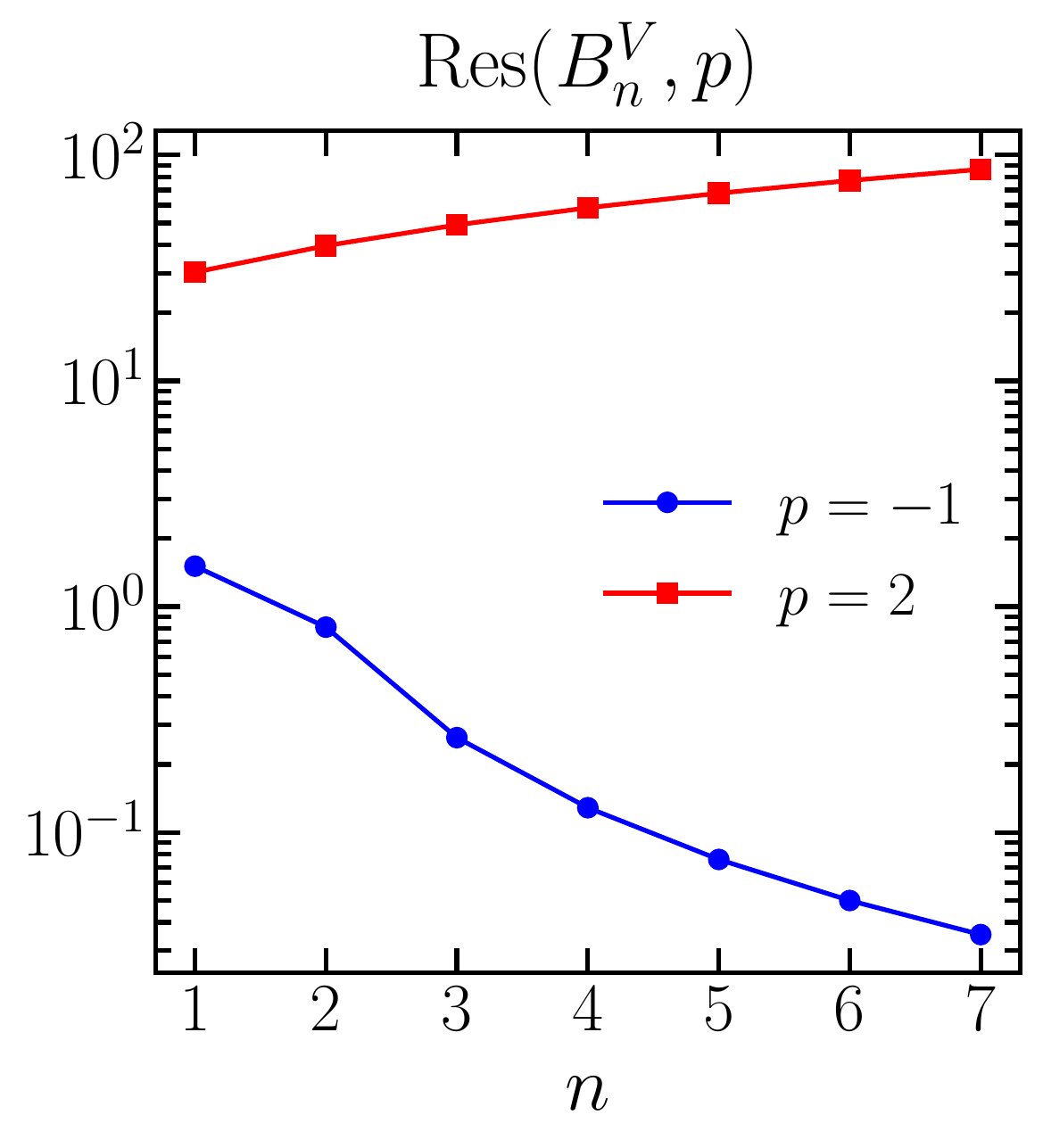}{\label{fig:Residues2}}}
\caption{(a) Absolute value of the residues of $B_{n}^V$ relative to those of $S_{n}^V$ for the leading UV and IR poles at $p=-1$ and $p=2$, respectively.
(b) Absolute value of the residues of $B_{n}^V$ for the same two poles. }
\label{Fig:Residues}
\end{center}
\end{figure}

The fact that $B_{n}^\delta$ is given by a difference of two Borel transforms suggests that renormalon cancellations may take place. We find that the residues of the leading UV and IR poles are significantly smaller in $B_{n}^V$ than their counterparts in $S_{n}^V$. For example, for the leading UV pole,
Fig.~\ref{fig:Residues1} shows that the residue at $u=-1$ of $B_{3}^V$ is $31(38)$ times smaller than that of $S_{3}^V$ ($S_{4}^V$). For the leading IR pole the residue of $B_{3}^V$ at $u=2$ is only $16.0\%(8.1\%)$ that of $S_{3}^V$ ($S_{4}^V$). Furthermore, in absolute terms, the residue of $R_{q,n}^V$ at $u=-1$ decreases as $n$ grows, as shown in Fig.~\ref{fig:Residues2}, which
leads to the expectation of an exact cancellation in the limit of $n\to \infty$. This can be corroborated by an analysis of the residue of the leading UV pole for large $n$. For the vector current one has that $P_n^V(-1)\simeq 0.7 n^{3/2}$, while the rest of terms in the residue at $u=-1$ tend to $6C_Fe^{-5/3}\sqrt{\pi/n}$ (with $\mu=\mbar_q$), such that the complete residue can be approximated by the linear expression $1.4 C_F n$. This, in turn, implies the conjectured cancellation and vector moment ratios have zero residue for $n\to \infty$ (decreasing as $1/n^2$). Very similar conclusions can be drawn for the pseudo-scalar correlator.

A similar observation can be made for the leading IR pole at $u=2$. This time, however, even though the dependence of the residue with $n$ is tamed for the moment ratios, it still grows (in absolute value) with $n$. This can be understood in the following way: the polynomials $P_n^V(u)$ evaluated at $u=2$ are all positive and grow approximately like $2.1 n^{7/2}$. At the sight of Eq.~\eqref{eq:asy} one concludes that the residue at $u=2$ in the $\MSb$ scheme is always positive and, given that the rest of terms behave as $(3/64)C_Fe^{10/3}\sqrt{\pi/n}$ for large $n$, can be approximated by $4.9C_F n^3$.
For moment ratios the residue becomes negative and softened to a linear expression: $-6.8 C_F (2.6 + n)$, as can be seen (in absolute value) in Fig.~\ref{fig:Residues2}.

We have also checked that if the quark mass is expressed in the pole scheme, the residues of the Borel transform $S_n^V$ at $u=2$ for the first four physical moments are significantly reduced.
When switching to the pole scheme one gets a negative contribution to the $u=2$ renormalon of the form $-n e^{10/3}C_F$ (common to all currents),
that is, proportional to $n$. This contribution is of similar size that the $\MSb$ term in absolute value for $n\leq3$, translating into a
significant cancellation (particularly strong for $n=1,2$). For larger values of $n$ the cancellation is less important, and becomes more and more
irrelevant as $n$ grows. (A similar behaviour is expected in the case of the axial-vector current.) This decrease of the $u=2$ residue should be regarded as accidental and not related to a softening of the non-perturbative contribution coming from the gluon condensate.

We observe that for the pseudo-scalar moments $M_{c,n}^P$, changing to the pole mass does not lead to a reduction in the residue of the leading
IR pole.
In this case, one has that $P_n^P(2)$ is negative for
$n<2$, positive for $n>2$, and, as already discussed, vanishes at $n=2$. Furthermore, they rapidly grow in absolute value as $n$ increases, and
one can conjecture again a $n^{7/2}$ behaviour. At the sight Eq.~\eqref{eq:asy} one easily sees that the non-polynomial terms yield a positive factor
that for large $n$ becomes again $(3/64)C_Fe^{10/3}\sqrt{\pi/n}$. Therefore one never has cancellations in this case because in the region where the
two contributions are of similar size they are both positive, and when signs become opposite the pole-mass correction is already much smaller than
the main term.

Since the leading renormalon singularities in $R_{q,n}^V$ are softened with respect to the moments $M_{q,n}^V$ we can expect that the perturbative behaviour of
the ratios should be significantly improved. In particular, the partial cancellation of the leading UV pole should lead to series that are better behaved,
specially for larger $n$, postponing the onset of the sign alternation pattern for the coefficients. For the moments and ratios of moments with $\delta=P$ a very similar scenario for the leading renormalon singularities emerges and we refrain from showing the equivalent of Fig.~\ref{Fig:Residues} for this case, but similar conclusions apply.\footnote{For the cases without a direct phenomenological application, namely $\delta=S$ and $A$, the singularities are again softened in the ratios of moments, but the cancellation of the leading UV renormalon when $n\to \infty$ is not apparent.} We investigate the perturbative expansion of the ratios with $\delta=P$ and $V$ in the light of our findings for the renormalons in
the next section.

\section{Perturbative expansion of \boldmath \texorpdfstring{$R_{q,n}^\delta$}{RnX} in the large-\texorpdfstring{$\beta_0$}{beta0} limit}
\label{sec:PtSeriesR}

\subsection{Higher order behavior of the perturbative series}

Let us turn to a study of the perturbative series of the moments $R_{q,n}^{\delta}$ with $\delta=V$, $P$ in the \Lb limit. The perturbative coefficients for the $\alpha_s$ expansions
of these ratios of moments can be obtained analytically from the expressions of $B_n^\delta(u)$ and the use of a formula analogous to
Eq.~\eqref{eq:CndeltaPtCoeff} but without the terms proportional to $\gamma_{m}^{(k)}$.
In the \Lb limit, the ``true value" for the moments $R_{q,n}^\delta$ is known and given by the Borel integral of Eq.~\eqref{eq:RnFullResult}, with an
imaginary ambiguity arising from the IR poles that is numerically quite small in our case. This result is scheme and scale independent, as already discussed. We restrict our analysis to ratios that involve moments with $n\leq 4$ because for larger $n$ the series is, effectively, an expansion in $\alpha_s\sqrt{n}$~\cite{Voloshin:1995sf} and we checked that our results have this behaviour for $n$ large, as expected.

\begin{figure}[!t]
\begin{center}
\subfigure[]{\includegraphics[width=0.32\textwidth]{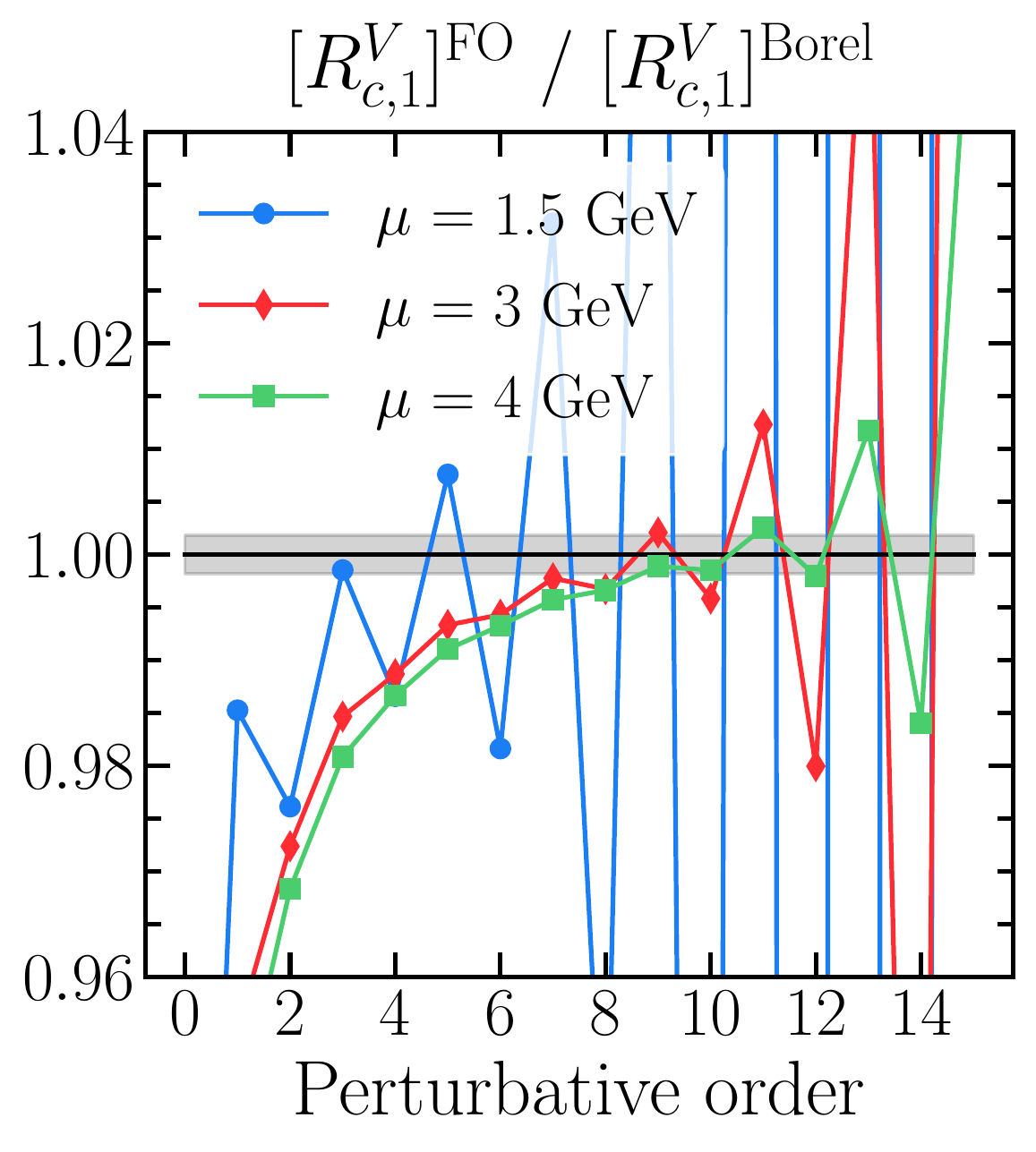}\label{Fig:CharmV_PertSeries_Largeb0n1}}
\subfigure[]{\includegraphics[width=0.32\textwidth]{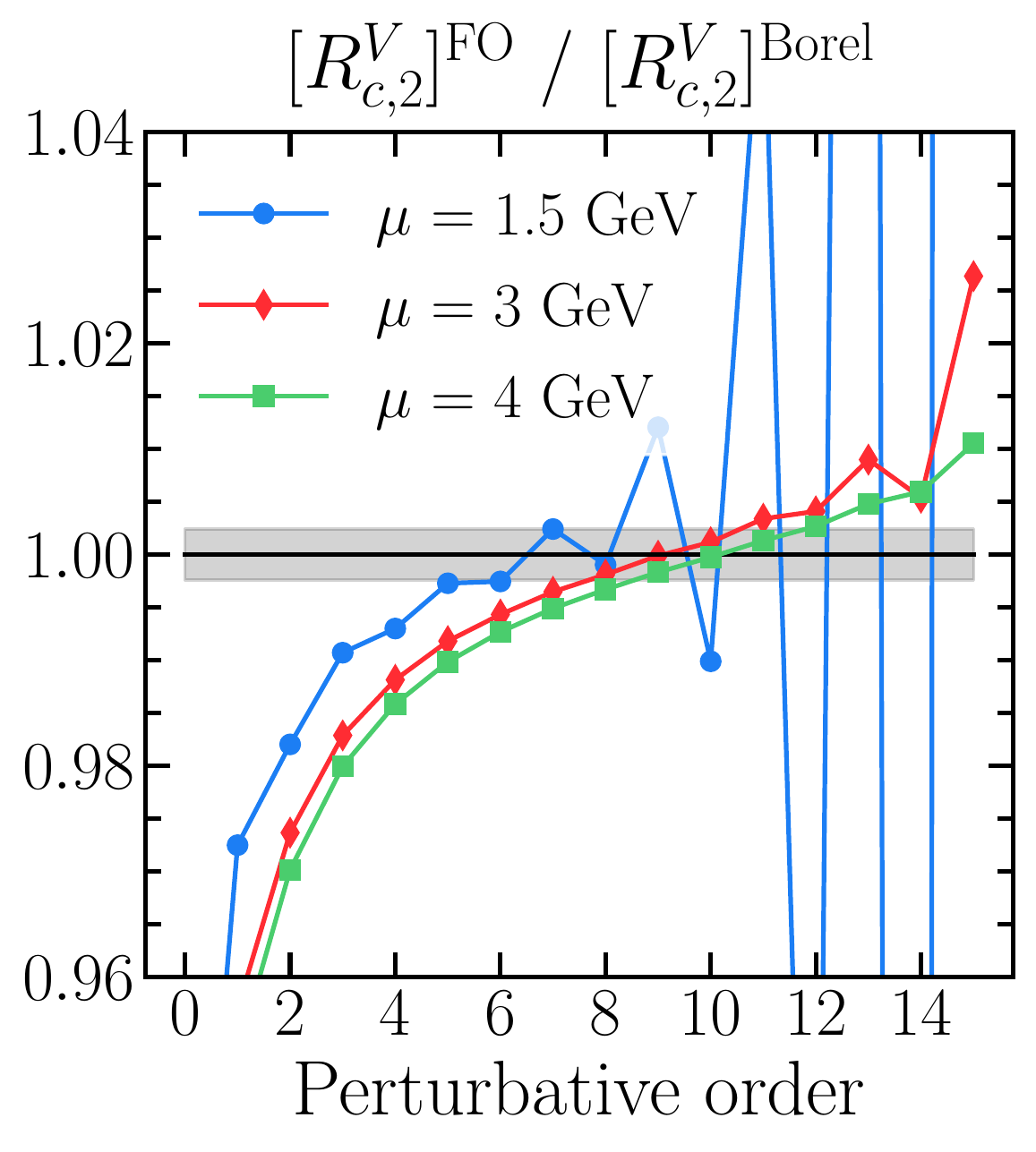}\label{Fig:CharmV_PertSeries_Largeb0n2}}
\subfigure[]{\includegraphics[width=0.32\textwidth]{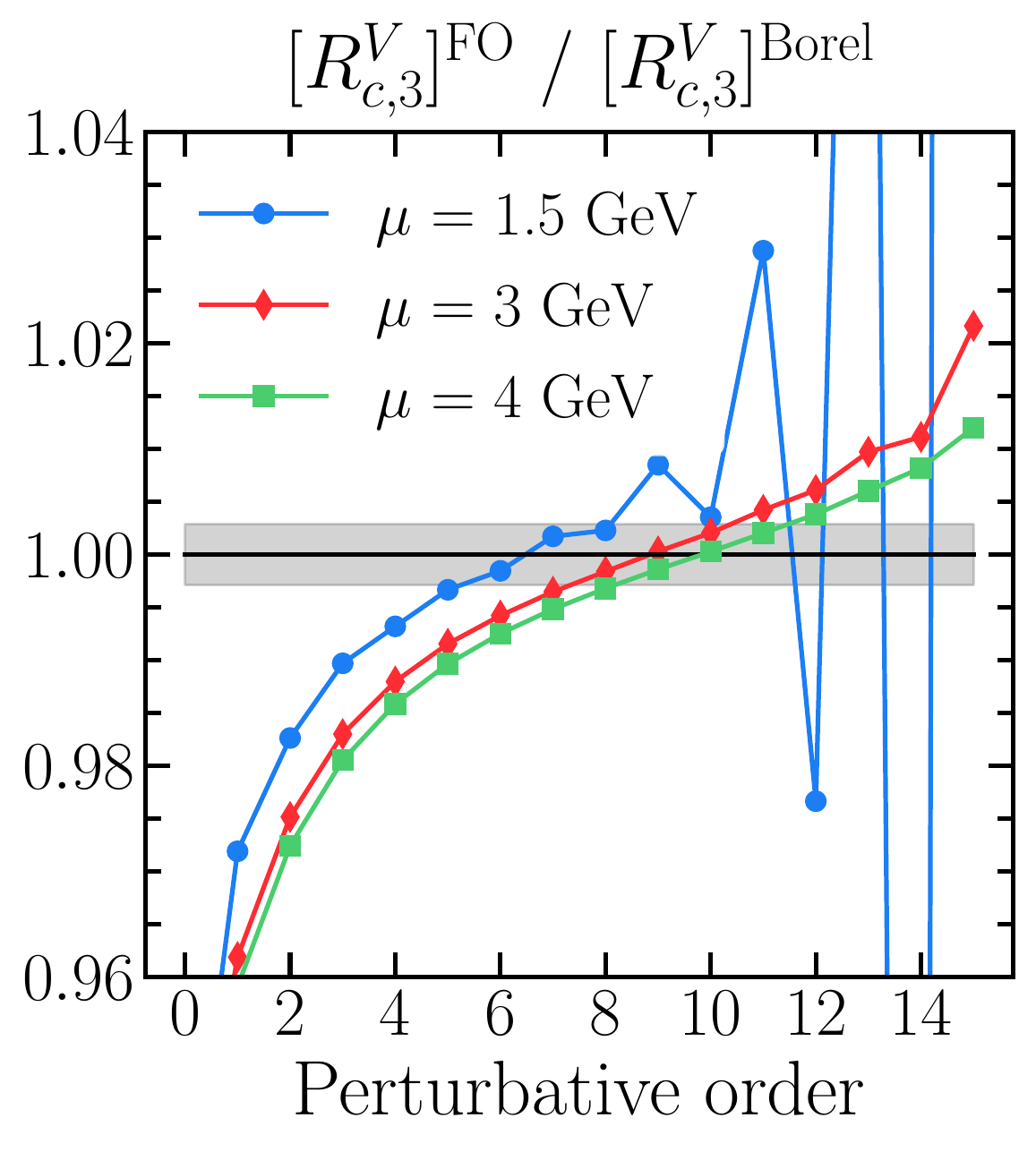}\label{Fig:CharmV_PertSeries_Largeb0n3}}
\caption{Perturbative series of ratios of vector-current charm moments $R_{c,n}^V$ in the large-$\beta_0$ limit normalized to the real part of the Borel integral of $R_{c,n}^V$. The gray band represents the ambiguity of the Borel integral.}
\label{Fig:CharmV_PertSeries_Largeb0All}
\end{center}
\end{figure}

In the perturbative expansion in powers of $\alpha_s(\mu)$ one has the usual freedom
of varying the renormalisation scale $\mu$, which is often used as a way to probe higher orders and assess the uncertainty associated with the truncation of
perturbation theory. One should recall that in these series, since the quark mass appears only in the argument of $1/\beta_0$-suppressed logarithms,
the running of the $\MSb$ quark mass, $\overline m_q(\mu)$, will generate $\mathcal{O}(1/\beta^2_0)$ or higher subleading terms. Therefore, here, we
will use the fixed reference masses $\mbar_c=1.28$\,GeV and $\mbar_b=4.18$\,GeV, which will not be RG-evolved in our
phenomenological explorations. For the strong coupling we
use the reference value $\alpha_s^{(n_f=5)}(m_Z) = 0.1179$, with $m_Z=91.1876\,$GeV~\cite{Zyla:2020zbs}, which yields $\alpha_s^{(n_f=4)}(\mbar_b)=0.2245$ and
$\alpha_s^{(n_f=3)}(\mbar_c)=0.3865$ using the five-loop running coupling~\cite{Baikov:2016tgj, Herzog:2017ohr,Luthe:2017ttg,Chetyrkin:2017bjc} and four-loop matching~\cite{Schroder:2005hy, Chetyrkin:2005ia} at the thresholds, both in full QCD. These values have been obtained with \texttt{REvolver}~\cite{Hoang:2021fhn}. The running of $\alpha_s(\mu)$ in the large-$\beta_0$ perturbative series is then performed at one-loop accuracy, for consistency. In this limit they correspond to $\Lambda_{\rm QCD}^{(n_f=4)}=145\,$MeV and $\Lambda_{\rm QCD}^{(n_f=3)}=210\,$MeV. In our large-$\beta_0$ analyses we will use $n_\ell=3$ and $n_\ell=4$ active flavors for charm and bottom moments, respectively.
The results obtained in this section were implemented in independent \texttt{Mathematica} and \texttt{Python} codes that agree to machine precision.

In Fig.~\ref{Fig:CharmV_PertSeries_Largeb0All} we show the perturbative expansion of the first three ratios $R_{c,n}^V$ for three choices of the
scale
$\mu$. In these plots, we normalize the results to the real part of the Borel integral of $R_{q,n}^V$ such that all the series should approach unity.
Since the UV poles lie at negative values of the variable $u$, their residues grow for lower renormalisation scales and it is expected that small
$\mu$ will enhance these singularities, as can be clearly
seen in Fig.~\ref{Fig:CharmV_PertSeries_Largeb0n1}, where the series with $\mu=1.5$\,GeV shows the sign alternation pattern typical of UV renormalons already
at the first few orders of perturbation theory. For larger $\mu$ the residue of the leading UV pole is smaller and this oscillation is postponed.
Ratios $R_{c,n}^V$ with higher values of $n$ have weaker UV renormalons, as shown in Figs.~\ref{Fig:CharmV_PertSeries_Largeb0n2}
and~\ref{Fig:CharmV_PertSeries_Largeb0n3},
which is a consequence of the partial cancellation of the leading UV pole discussed in the previous section. However, the series for higher $n$ do
not stabilize around the true value given by the Borel integral. Instead, they cross this value with a fixed sign pattern and later run into the
asymptotic regime. This is typical of series that have a large IR renormalon~\cite{Beneke:2012vb,Boito:2020hvu} and is in full agreement with the
discussion in Sec.~\ref{eq:ratiosLarge}, namely that the residue of the leading IR pole grows with $n$. Another salient feature of these results is that the partial cancellation of the
leading UV renormalon leads to series that are somewhat better behaved but that do not necessarily approach the true value faster. In fact, it turns
out that for higher $n$ the series truncated at $\alpha_s^3$ are further away from the true result.

\begin{figure}[!t]
\begin{center}
\subfigure[]{\includegraphics[width=0.32\textwidth]{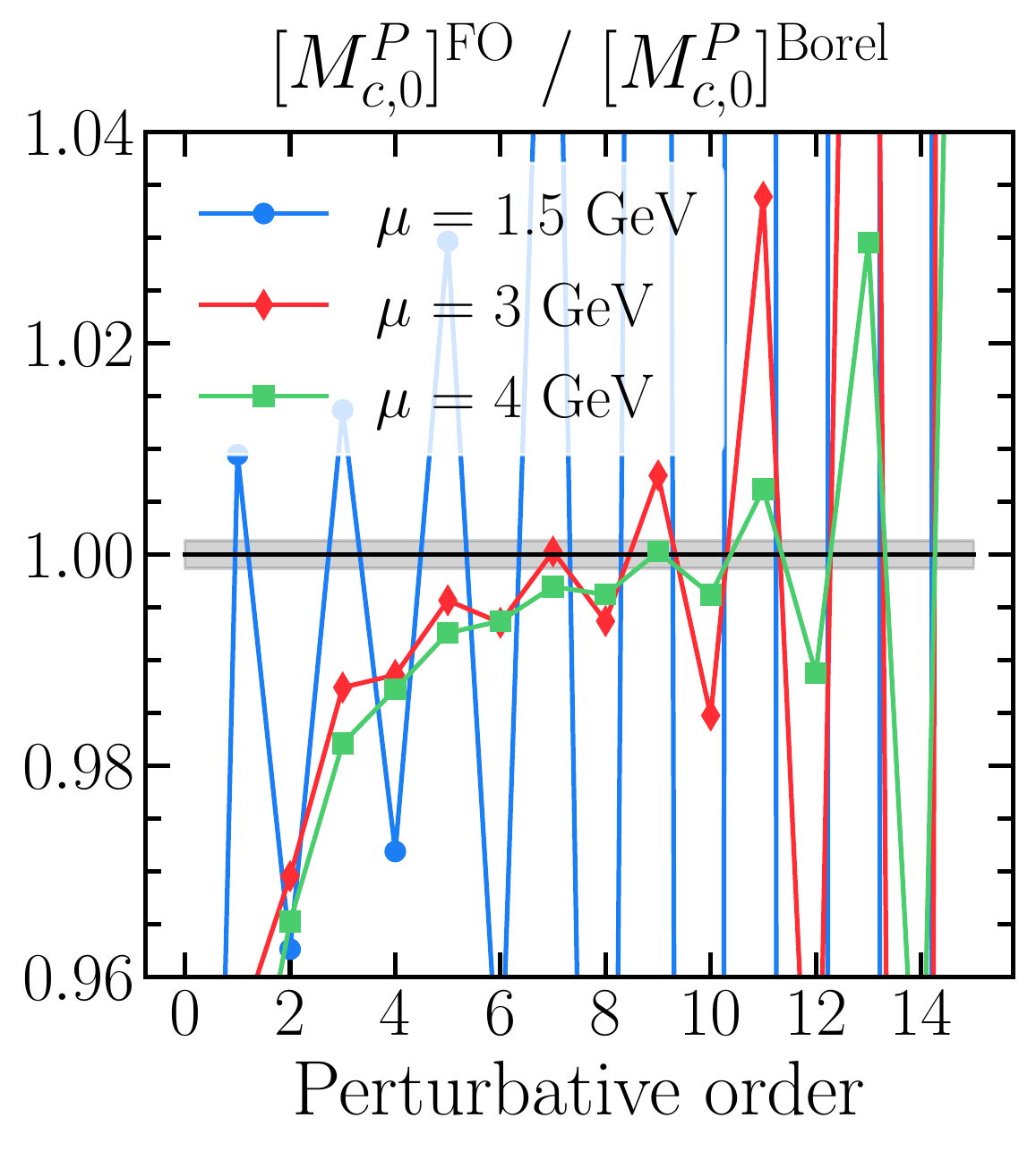}\label{Fig:CharmP_PertSeries_Largeb0n1}}
\subfigure[]{\includegraphics[width=0.32\textwidth]{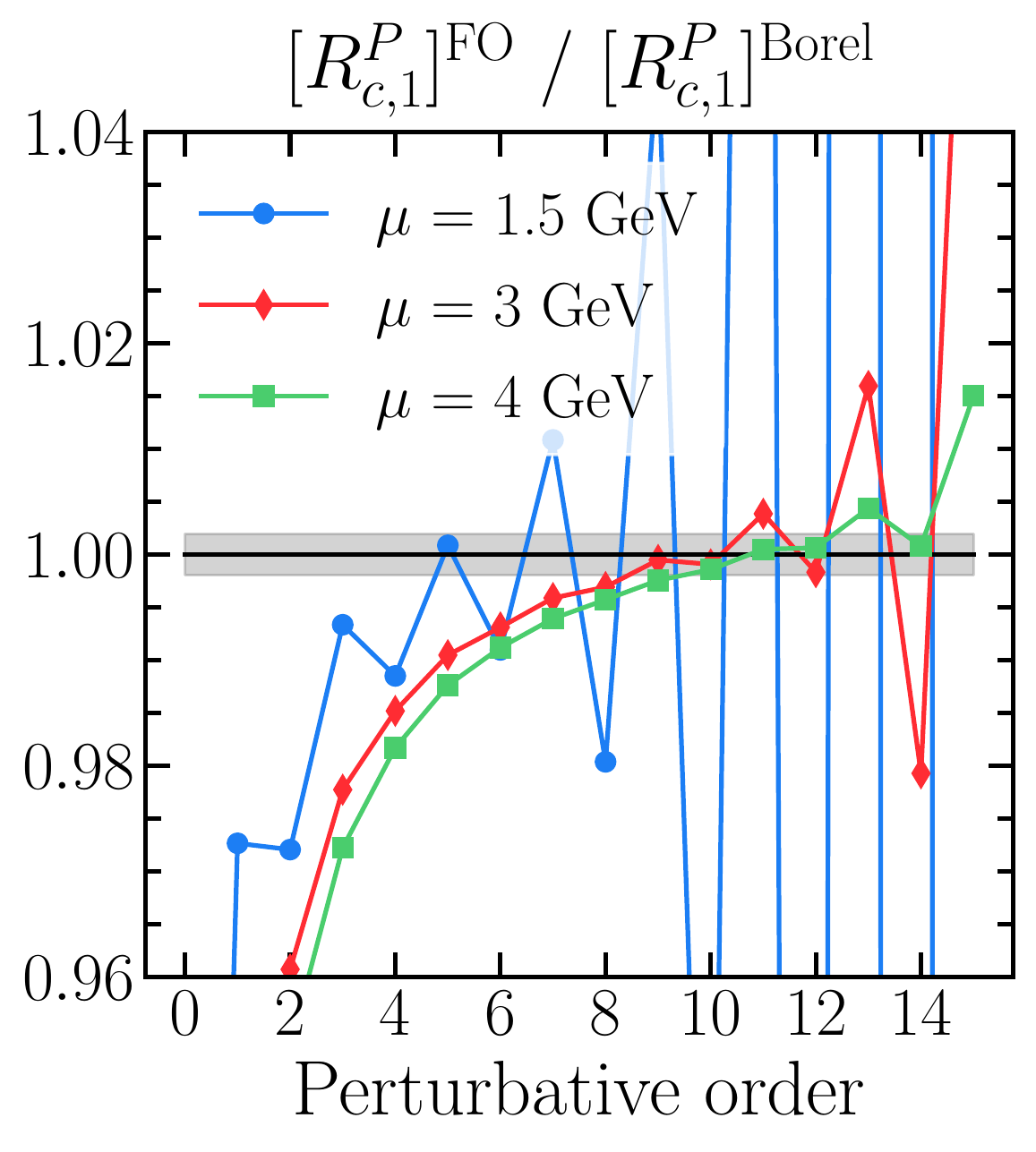}\label{Fig:CharmP_PertSeries_Largeb0n2}}
\subfigure[]{\includegraphics[width=0.32\textwidth]{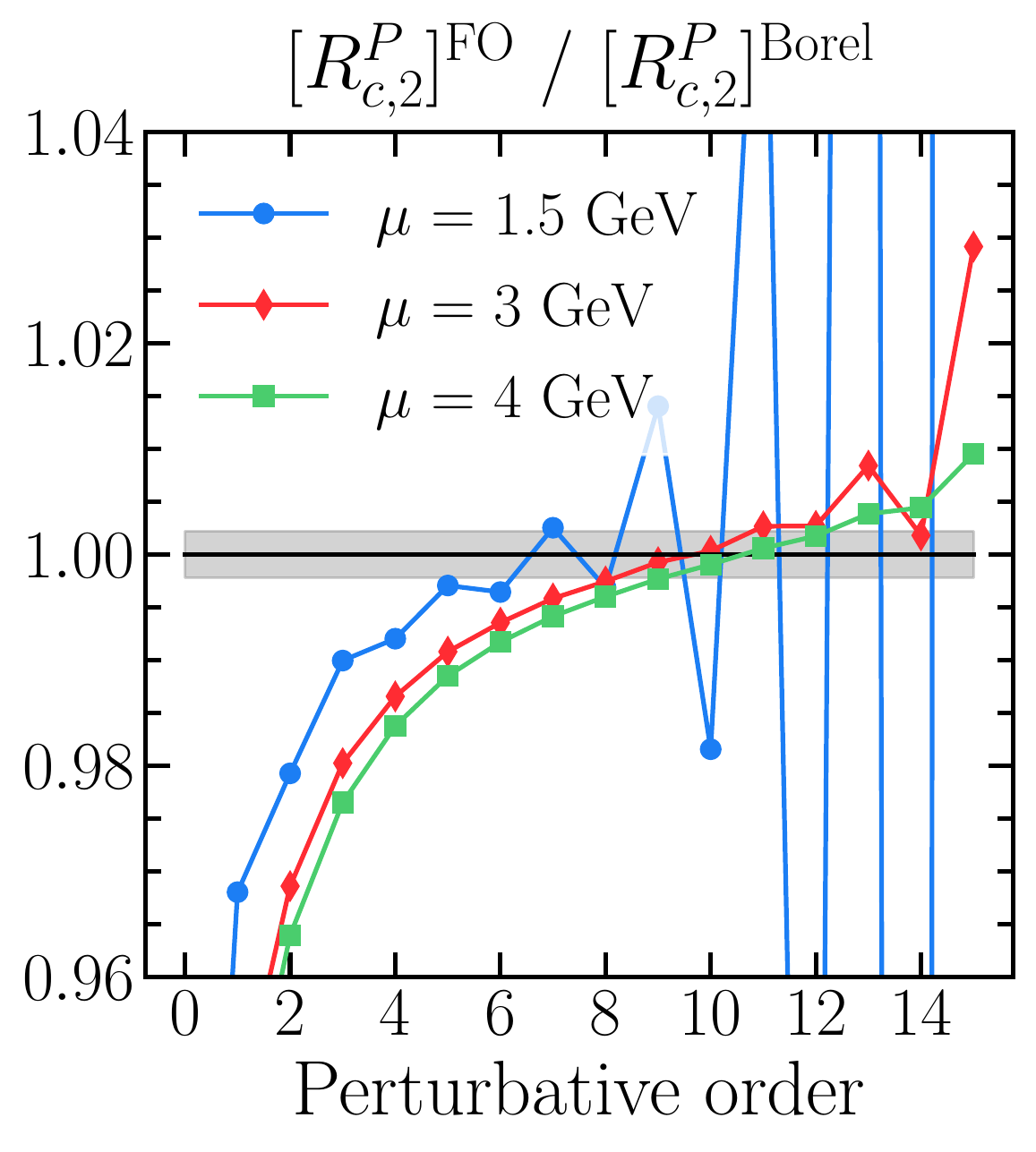}\label{Fig:CharmP_PertSeries_Largeb0n3}}
\caption{Perturbative series of ratios of pseudo-scalar current charm moments $R_{c,n}^P$ in the large-$\beta_0$ limit normalized to the real part of the Borel integral of $R_{c,n}^P$. The gray band represents the ambiguity of the Borel integral.\label{Fig:CharmP_PertSeries_Largeb0All}}
\end{center}
\end{figure}

In Fig.~\ref{Fig:CharmP_PertSeries_Largeb0All} we show similar results for the pseudo-scalar current correlator. Here, we start with the moment $M_{c,0}^P$ which does not have the mass dependent pre-factor and is therefore a quantity completely analogous to the ratios $R_{c,n}^P$. However,
this moment cannot benefit from the partial cancellation of renormalons that we discussed in the previous section, since its Borel transform is given
solely by Eq.~\eqref{eq:SnP}. We see in Fig.~\ref{Fig:CharmP_PertSeries_Largeb0n1} that this moment has a very large contribution of the UV
singularities, with sign alternation clearly visible even for high values of $\mu$. For the ratio $R_{c,1}^P$, we see in
Fig.~\ref{Fig:CharmP_PertSeries_Largeb0n2} that the partial cancellation is now in place, but the sign alternation is still present at lower orders
and only for $R_{c,2}^V$ this behaviour starts to be tamed.

\begin{figure}[!t]
\begin{center}
\subfigure[]{\includegraphics[width=0.32\textwidth]{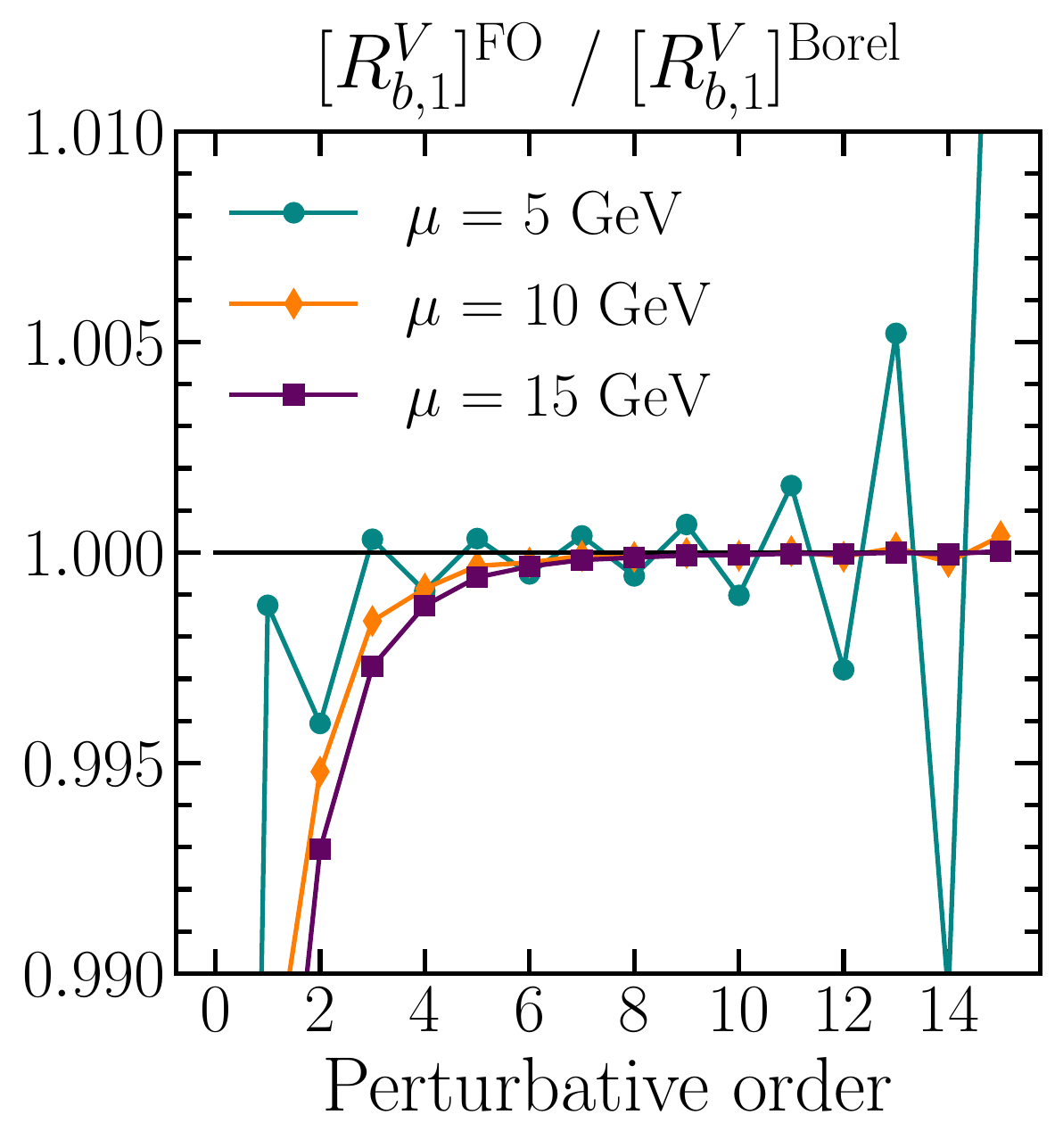}\label{Fig:BottomV_PertSeries_Largeb0n1}}
\subfigure[]{\includegraphics[width=0.32\textwidth]{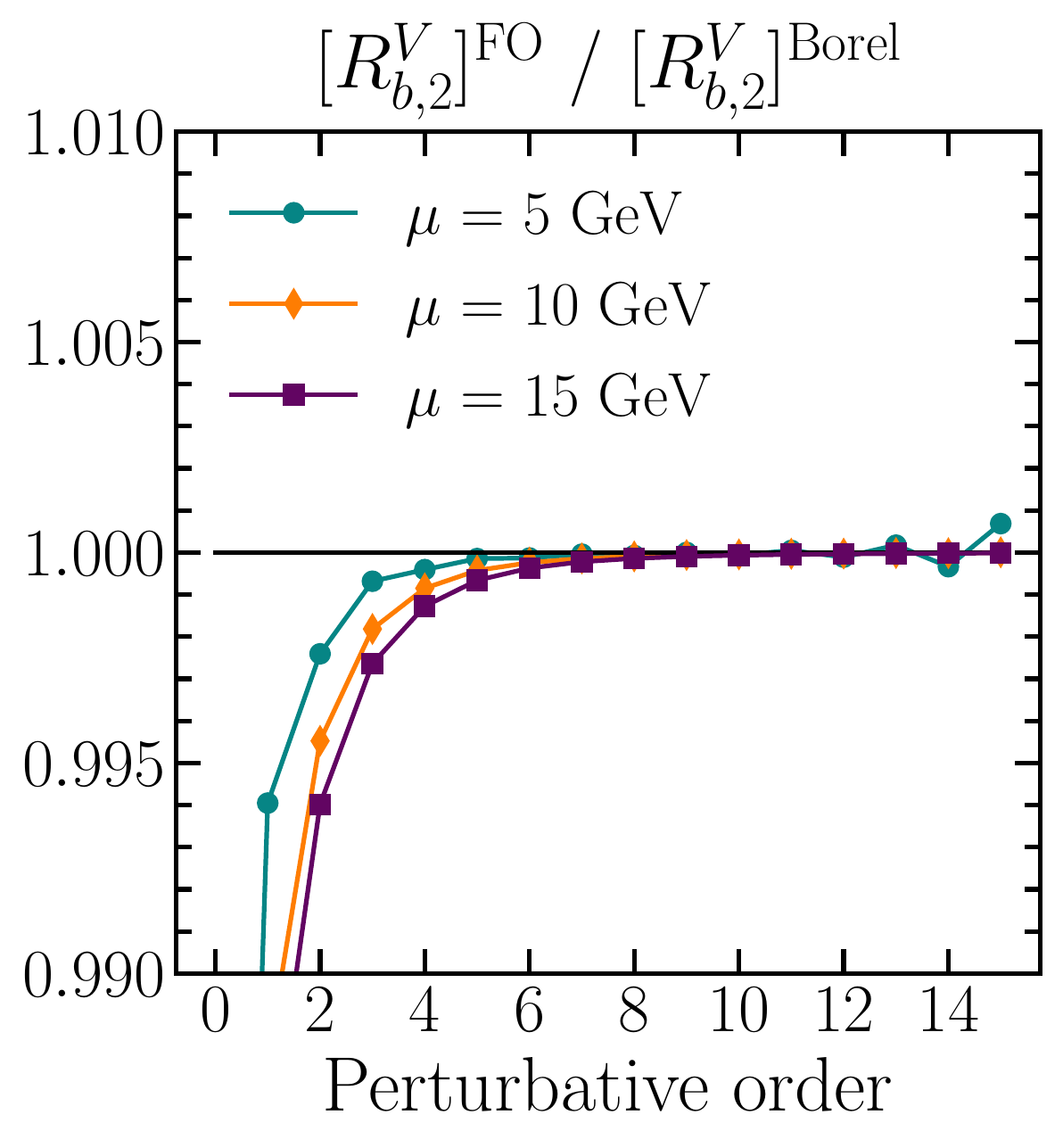}\label{Fig:BottomV_PertSeries_Largeb0n2}}
\subfigure[]{\includegraphics[width=0.32\textwidth]{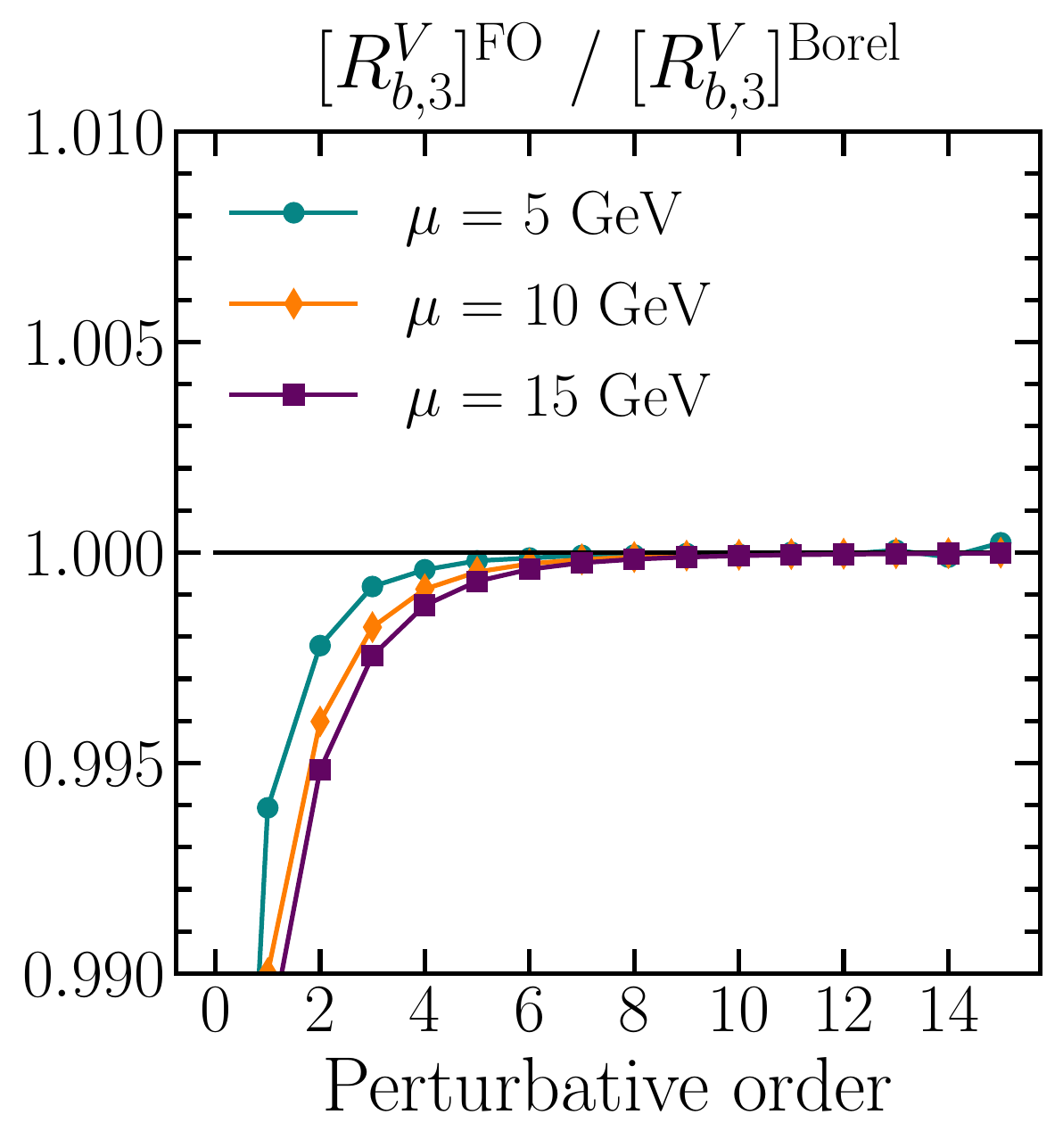}\label{Fig:BottomV_PertSeries_Largeb0n3}}
\caption{Perturbative series of ratios of vector-current bottom moments $R_{b,n}^V$ in the large-$\beta_0$ limit normalized to the real part of the Borel integral of $R_{b,n}^V$.}
\label{Fig:BottomV_PertSeries_Largeb0All}
\end{center}
\end{figure}

Finally, in Fig.~\ref{Fig:BottomV_PertSeries_Largeb0All} we show the results for the first three bottom-quark vector correlator ratios $R_{b,n}^V$.
The main difference in this case is that, overall, all the series are much better behaved, which simply reflects the fact that $\alpha_s(\mu)$ is now
much smaller, postponing the onset of the asymptotic regime to significantly higher orders.
Again, for $n=1$ with the lowest value of scale, here
$\mu=5$\,GeV, the effects of the leading UV pole are clearly seen in the sign alternation of the series coefficients. For higher values of $n$, the
partial cancellation of the UV renormalon leads to series with a uniform approach to the true value. Albeit very well behaved, all the series
approach the true value somewhat slowly, and at $\mathcal{O}(\alpha_s^3)$ a relatively large spread with scale variation is still visible.
The ambiguity arising from IR poles is tiny and not visible in the plots of Fig.~\ref{Fig:BottomV_PertSeries_Largeb0All}. This reinforces that non-perturbative effects are negligible in the vector bottom ratios $R_{b,n}^V$.

\subsection[From the large-\texorpdfstring{$\beta_0$}{beta0} limit to QCD]{From the \boldmath large-\texorpdfstring{$\beta_0$}{beta0} limit to QCD}

Before we can use the \Lb results for $R_{q,n}^\delta$ to derive consequences for their counterpart in QCD, it is important to compare the results
up to $\mathcal{O}(\alpha_s^3)$, the last order known in QCD, and assess how close the two series are. Our goal here is not to use the \Lb results as
an estimate of the unkown higher-order coefficients. Rather, we intend to derive more general conclusions that could guide the phenomenological applications, with special focus on the ratios $R_{q,n}^\delta$ with $\delta=V,$ $P$. In full QCD, we use four and five active flavors for charm and bottom moments, respectively. Furthermore, to mimic as much as possible our
large-$\beta_0$ analyses, we set $\mu_m=\mbar_q$ and identify $\mu_\alpha=\mu$ in Eq.~\eqref{eq:RqnPTExp}.

\begin{figure}[!t]
\begin{center}
\subfigure[]{\includegraphics[width=0.32\textwidth]{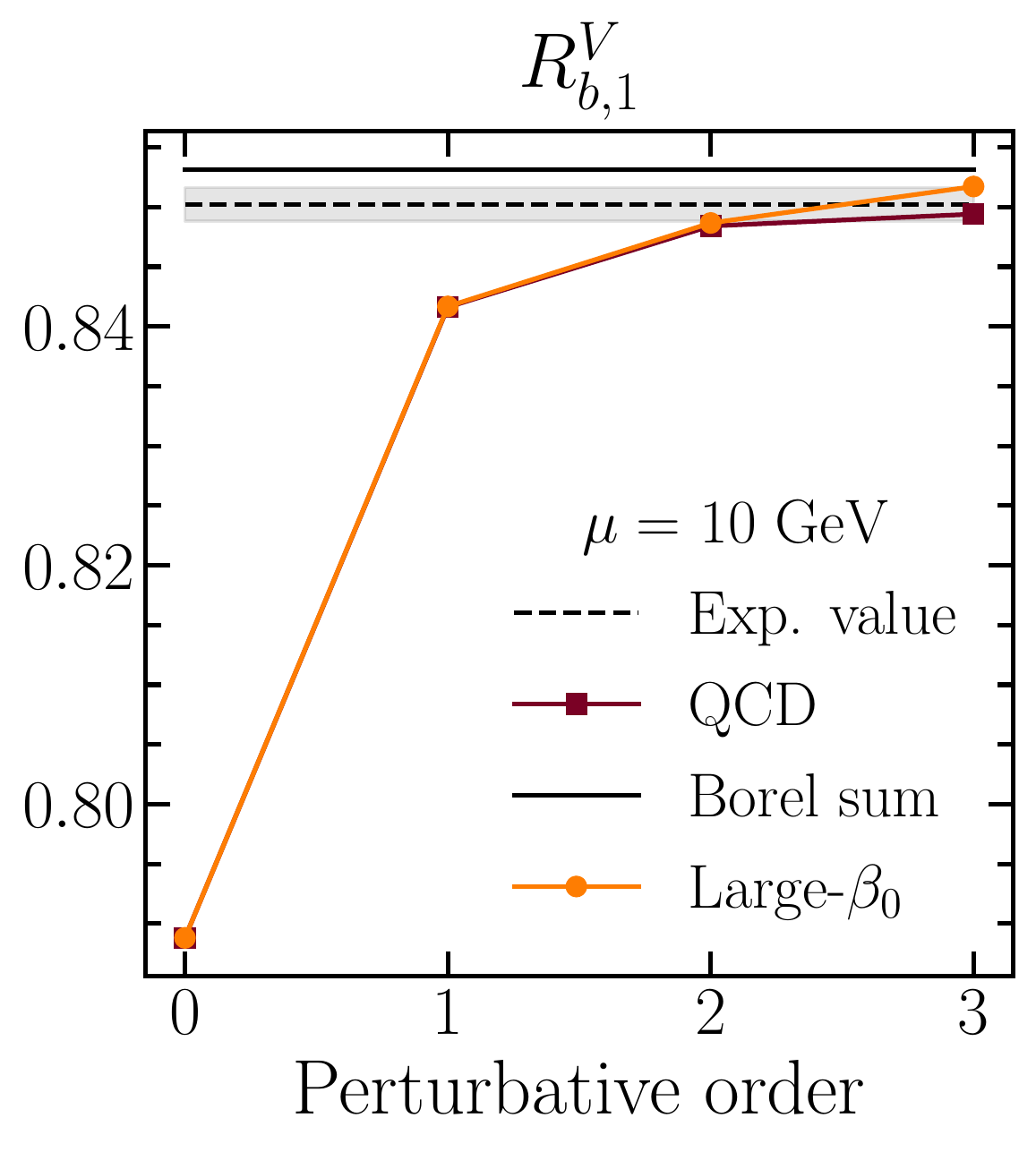}\label{Fig:QCDLbComparisonHigh(a)}}
\subfigure[]{\includegraphics[width=0.32\textwidth]{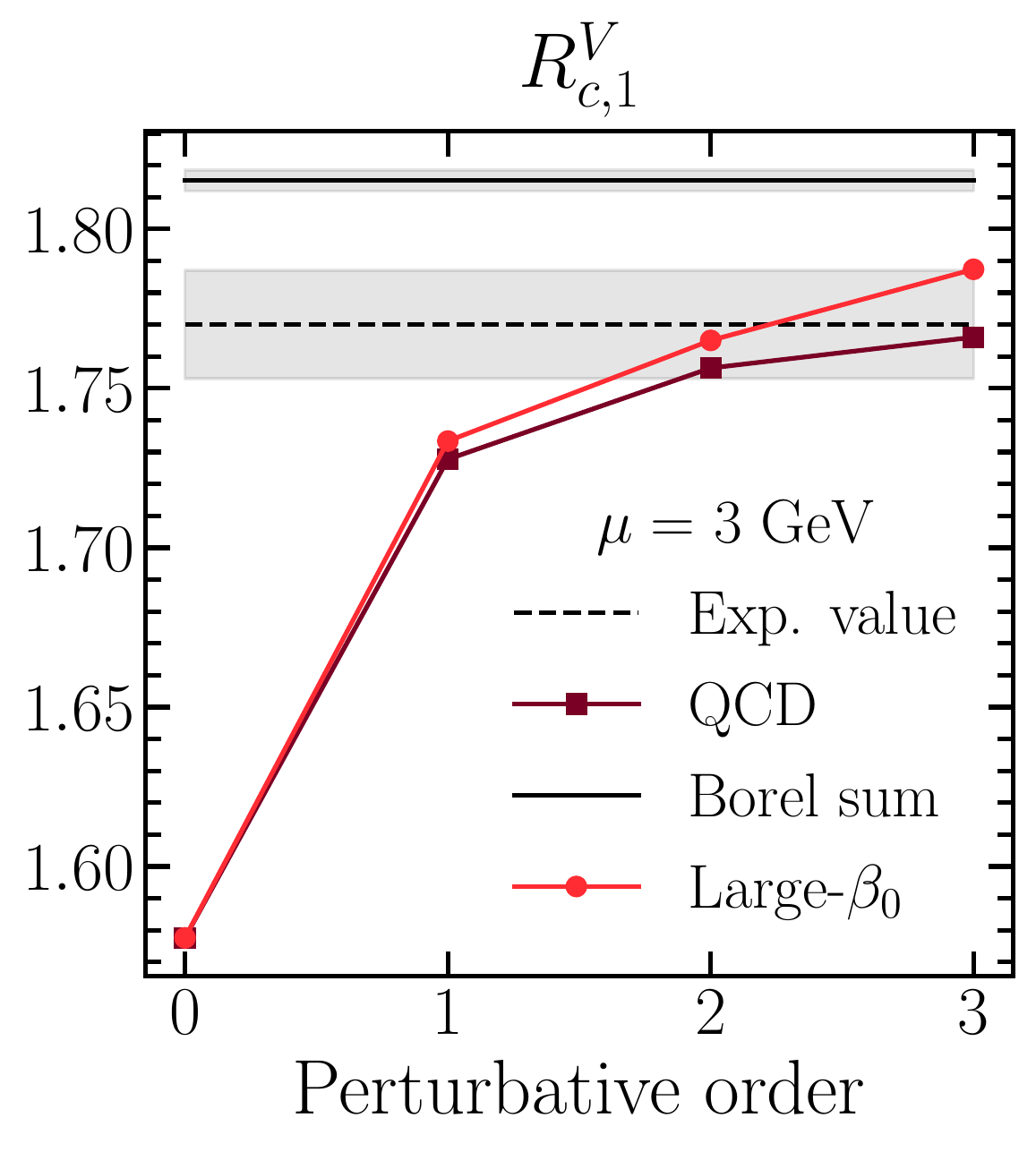}\label{Fig:QCDLbComparisonHigh(b)}}
\subfigure[]{\includegraphics[width=0.32\textwidth]{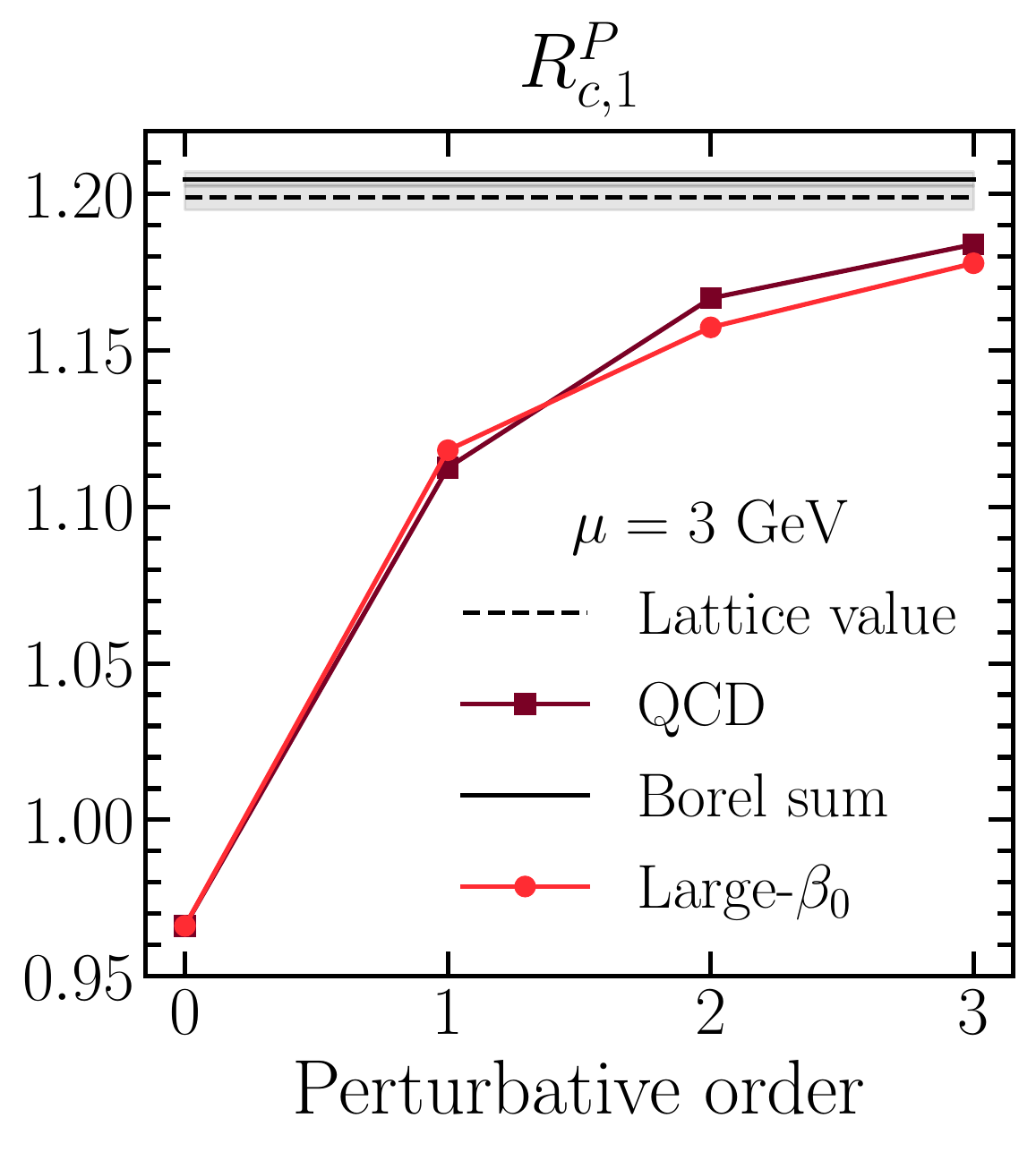}\label{Fig:QCDLbComparisonHigh(c)}}

\subfigure[]{\includegraphics[width=0.32\textwidth]{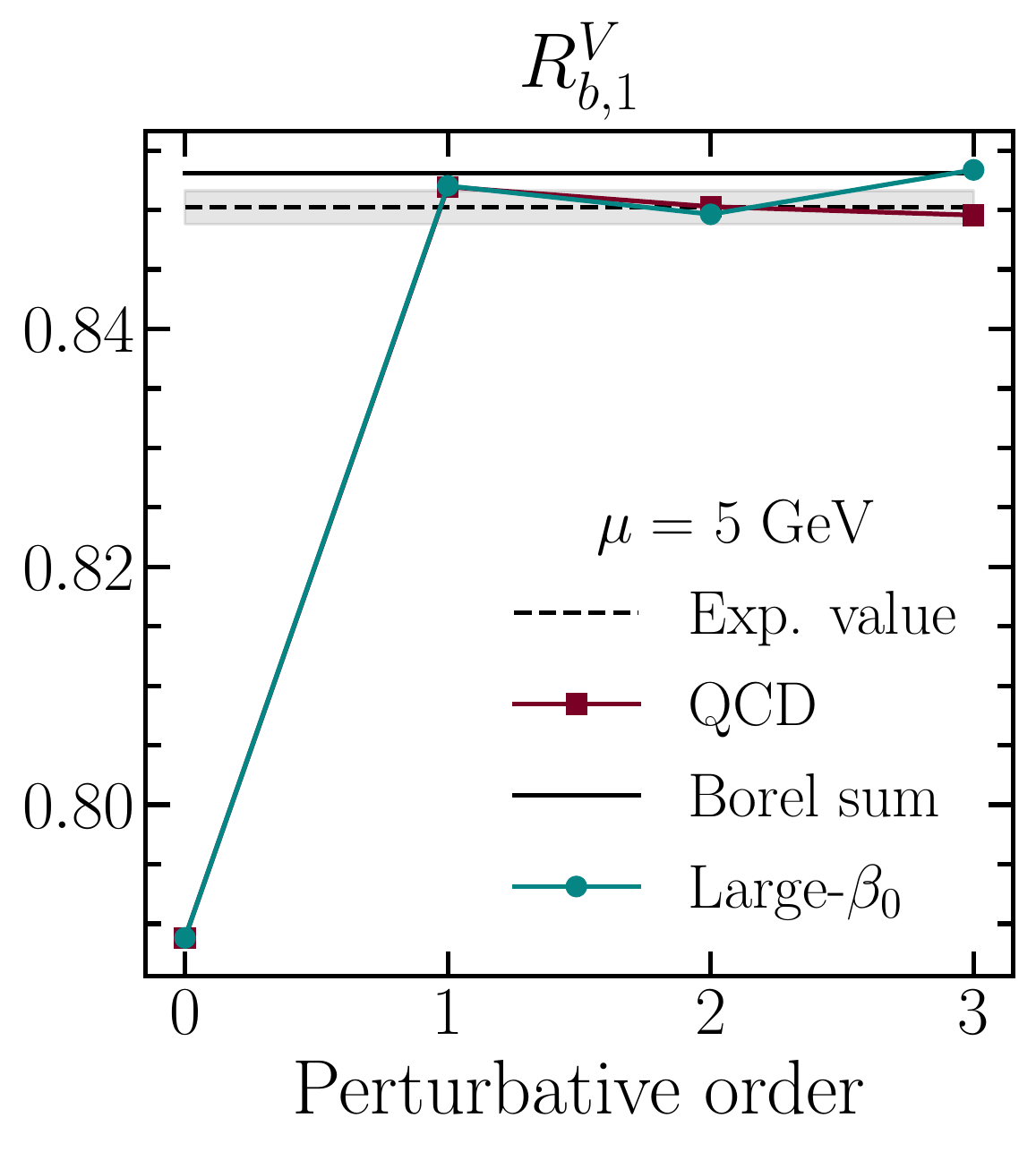}\label{Fig:QCDLbComparison(a)}}
\subfigure[]{\includegraphics[width=0.32\textwidth]{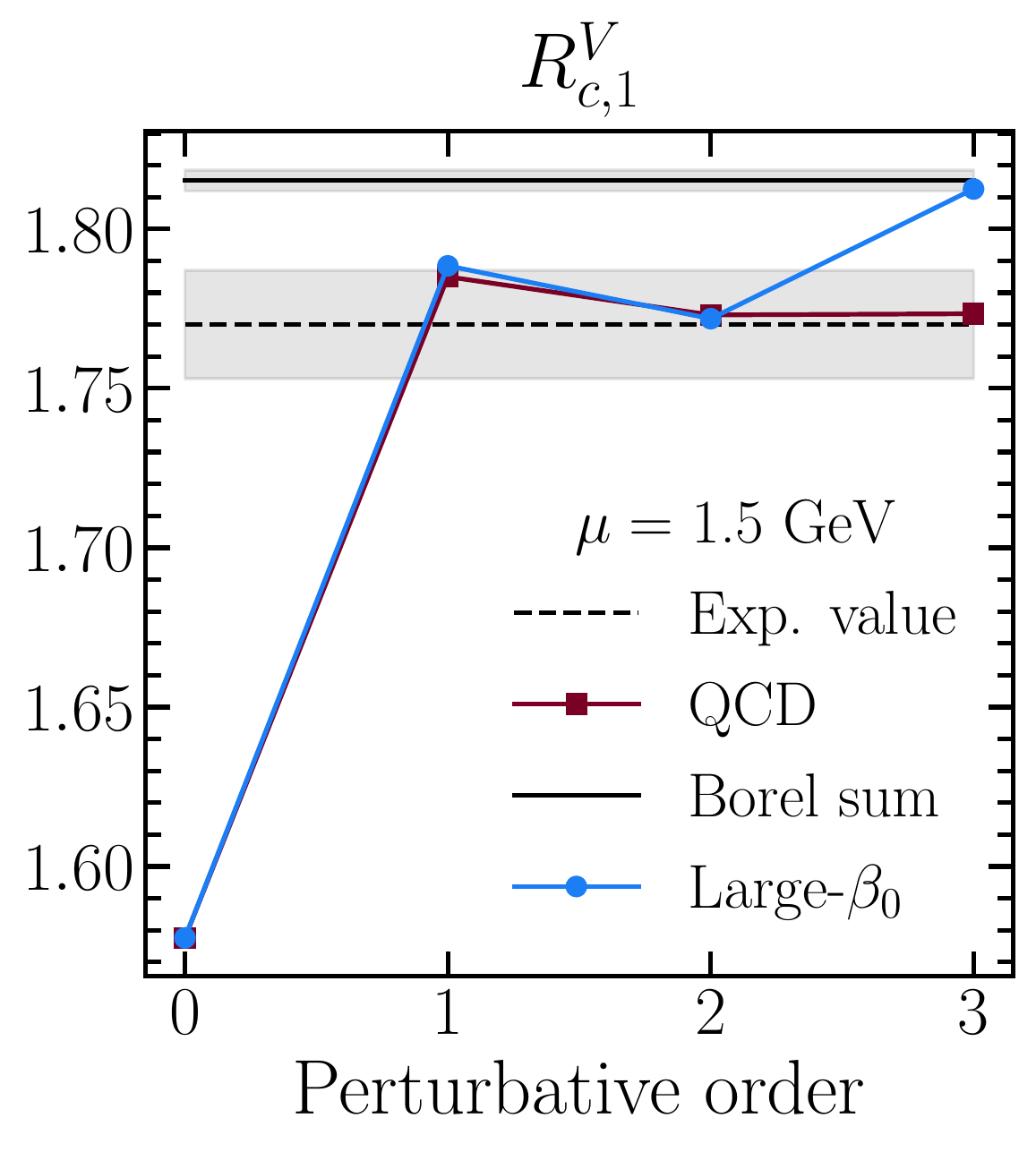}\label{Fig:QCDLbComparison(b)}}
\subfigure[]{\includegraphics[width=0.32\textwidth]{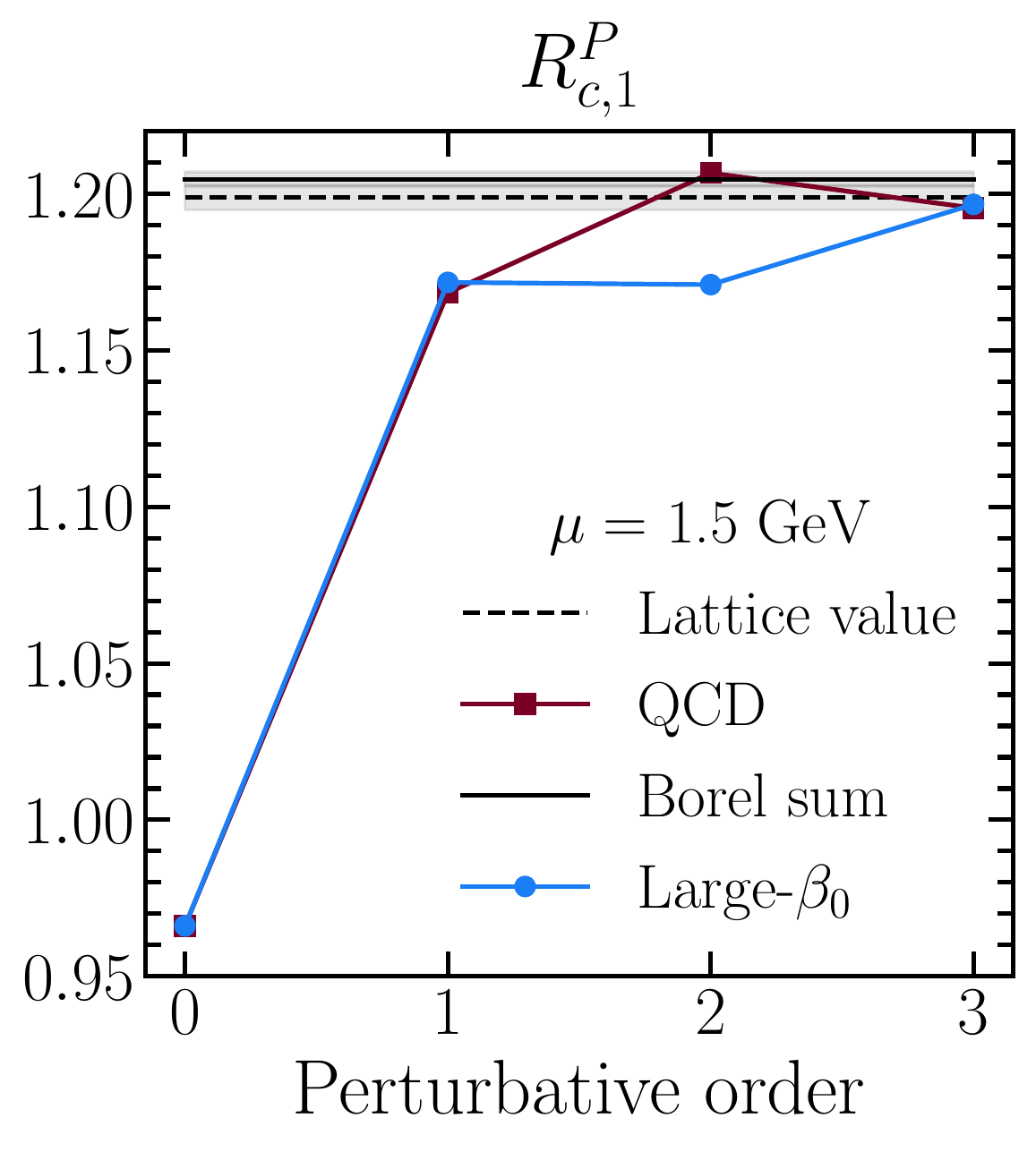}\label{Fig:QCDLbComparison(c)}}
\caption{Perturbative series for exemplary ratios of moments in \Lb and full QCD for $\mu\sim 2\mbar_q$ (upper panels) and $\mu\sim \mbar_q$ (lower panels). The solid horizontal line represents the \Lb Borel sum while the dashed lines are the values of the ratios obtained from experimental data $R_{b,1}^V = 0.8502\pm 0.0014$, $R_{c,1}^V = 1.770\pm0.017$~\cite{Boito:2020lyp,Boito:2019pqp} or lattice simulations $R_{c,1}^P = 1.199\pm 0.004$~\cite{Maezawa:2016vgv}.}
\label{Fig:QCDLbComparison}
\end{center}
\end{figure}

Let us start with a direct comparison of the series obtained in \Lb and QCD for three exemplary ratios of moments with $\mu\sim 2\mbar_q$. We see in the upper panels of
Fig.~\ref{Fig:QCDLbComparison} that the \Lb results do capture most of the features of the QCD series. There is, however, a difference related to the leading UV renormalon.
As we have shown, in \Lb, for lower renormalisation scales the dominance of the UV singularity is established at very low orders, which is
manifest in the sign alternation of the perturbative series coefficients, defined in Eq.~(\ref{eq:bar_rCoeff}), which produces a large order-by-order oscillatory behavior in the associated partial sum. In QCD, lowering the
renormalisation scale does not produce the same effect. Some of the coefficients do change sign, but no
systematic sign alternation emerges, as can be seen in the lower panels of Fig.~\ref{Fig:QCDLbComparison}. In particular in
panels~\ref{Fig:QCDLbComparison(a)} and~\ref{Fig:QCDLbComparison(b)}, the coefficients flip sign at
$\mathcal{O}(\alpha_s^2)$ but in QCD the coefficient remains negative for $\alpha_s^3$ corrections
as well. This means that the UV renormalon is not as salient
as in \Lb and that, likely, a competition
between IR and UV renormalons persists at intermediate orders even for
significantly low renormalisation scales. (This has already been observed in the context of the Adler
function~\cite{Boito:2018rwt}.) Therefore, the series coefficients at low renormalisation scales can be
significantly different between
\Lb and QCD. In particular, the independent coefficients $r_{i,0,0}^{\delta,(n)}$ of
Eq.~\eqref{eq:RqnPTExp} are not well reproduced beyond
$\alpha_s^2$, since they are evaluated at $\mu=\overline m_q$.
However, for larger renormalisation scales, for which the dominance of the UV pole has already subsided
in \Lb, the series can be quite similar to full QCD up to $\mathcal{O}(\alpha_s^3)$.

Another general observation of Fig.~\ref{Fig:QCDLbComparison}
is that, fortunately, the QCD series appear to approach
the data-based determinations of the ratios of moments faster than the series in \Lb approach the Borel sum. We also remark that the Borel sum in \Lb is in very good agreement with the data-based determination of the ratios of vector moments as well as the lattice determination of the pseudo-scalar ratios of moments with, perhaps, the exception of the ratios $R_{c,n}^V$, but even those are still marginally compatible since they have larger uncertainties.

\subsection{Combined ratios of moments}
\label{sec:NewRatios}

With the knowledge of the renormalon structure of the ratios of moments in \Lb we can construct new dimensionless combinations designed to further suppress or even exactly cancel specific renormalons. Ideally, one should rely on combinations that involve, at most, the first four physical moments, since these are well described within pQCD. From the vector moments with $n\leq 4$ and pseudo-scalar moments with $0\leq n\leq 3$, general dimensionless combinations are given by\footnote{In principle, one can even consider combinations of vector and pseudo-scalar moments, but we do not explore this possibility.}
\begin{align}
\widehat{R}^V_{q}(a, b, c) \equiv\,& [R_{q,1}^V]^a [R_{q,2}^V]^b [R_{q,3}^V]^c\,,\nn\\
\widehat{R}^P_{q}(a, b, c) \equiv\,& [M_{q,0}^P]^a [R_{q,1}^P]^b [R_{q,2}^P]^c\,,
\end{align}
with arbitrary real parameters $a$, $b$, and $c$.
The large-$\beta_0$ limit of $\widehat{R}^\delta_q$ is obtained by consistently re-expanding in $1/\beta_0$ the given combination using the results of Eq.~\eqref{eq:RnFullResult}. The Borel transform of $\widehat{R}^\delta_n$ can then be easily written in terms of $B_n^\delta(u)$.

The numerators of the leading IR and UV renormalons now become linear combinations of the parameters $a,b,c$. Suitable choices of these values can lead to significant reductions
of renormalon contributions to the perturbative series. Reducing the contribution from the IR renormalon at $u=2$ is of particular importance for charm correlators since
it is responsible for the runaway behaviour observed in the charm ratios $R_{c,n}^\delta$ displayed in
Figs.~\ref{Fig:CharmV_PertSeries_Largeb0All} and~\ref{Fig:CharmP_PertSeries_Largeb0All}, as well as being directly connected with the non-perturbative contribution from the gluon condensate. However, working with a combination that makes the $u=2$ residue vanish can lead to an enhancement of the $u=-1$ pole and accordingly to perturbative series highly dominated by
the leading UV renormalon with a sign-alternating behaviour already at low orders, even for high values of $\mu$. Therefore, one must achieve some compromise between the suppression of the leading IR singularity and the enhancement of the leading UV. For bottom ratios, given the tiny impact of the gluon condensate, finding a combination with no $u=-1$ singularity seems the best strategy.
Given that for $n=1$ such pole is double, the combination should be restricted to $n=2,3$ (that is, with $a=0$). But since $\alpha_s$ determinations from bottom moments are, at present, severely afflicted by large experimental errors we
do not explore this possibility any further.

\begin{figure}[!t]
\begin{center}
{\includegraphics[width=0.35\textwidth]{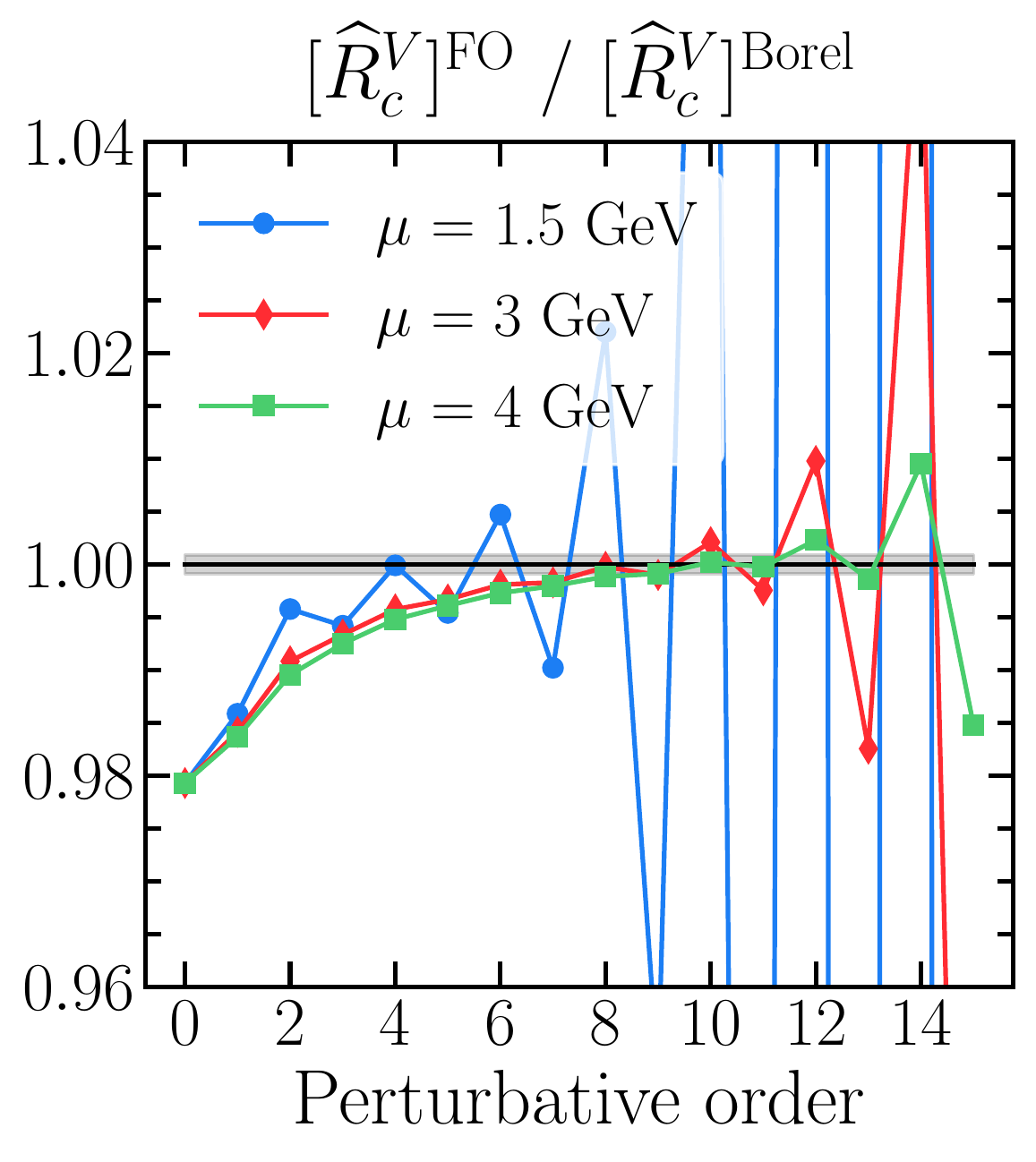}}
\caption{Perturbative series of the combination $\widehat{R}_c^V(-1/3, 1, -1/3)$ in large-$\beta_0$ normalised to the real part of its Borel integral. The gray band represents the ambiguity of the Borel integral.}
\label{Fig:CombinedRatioCharm}
\end{center}
\end{figure}

For illustration purposes we show in Fig.~\ref{Fig:CombinedRatioCharm} the perturbative expansion of the combined charm vector ratio
$\widehat{R}_c^V(-1/3,1,-1/3)$ for three values of $\mu$. This choice for the parameters reduces both the leading IR and UV residues by about
$70\%$, while the double UV pole present in $R_{c,1}^V$ is suppressed only by the value of the parameter $a$. A competition between both renormalons remains such that the perturbative series is not fully dominated by a fixed-sign or a sign-alternating behavior.
When compared to $R_{c,2}^V$ shown in Fig.~\ref{Fig:CharmV_PertSeries_Largeb0n2}, from which the main results of
Refs.~\cite{Boito:2019pqp, Boito:2020lyp} are based, we see that the perturbative series of $\widehat{R}_c^V(-1/3, 1, -1/3)$ approaches faster the
true value given by the Borel sum, has a weaker dependence on the renormalisation scale, and does not present a run-away behaviour typically seen in series dominated by IR renormalons. We have also checked from a direct comparison that the large-$\beta_0$ series of $\widehat{R}_c^V(-1/3, 1, -1/3)$ reproduces the non-log coefficients predicted by its QCD counterpart with great precision, and thus the large-$\beta_0$ series captures the features of the QCD series even at low values of $\mu$.

\subsection{Discussion}
With the above observations we are in a position to draw a few conclusions and advance a number of plausible hypothesis about the results in QCD and the impact on $\alpha_s$ extractions. Concerning the ratios $R_{q,n}^\delta$ with $\delta=V,$ $P$:
\begin{itemize}
\item We demonstrated that, in \Lb, these ratios benefit from partial cancellation of the leading renormalons. In the case of QCD,
the renormalon singularities become branch cuts, but the similarities between the results in \Lb and QCD allow us to speculate that an analogous
mechanism for the softening of the singularities seems to be at work for QCD as well. This strengthens the case for the use of these ratios in
$\alpha_s$ extractions.
\item In \Lb, the perturbative series for the ratios with $\delta=V$, $P$ are well behaved for not too low $\mu$. However, a relatively large spread arising from scale variations remains at $\mathcal{O}(\alpha_s^3)$. This spread is significantly reduced at $\mathcal{O}(\alpha_s^4)$ which indicates that the perturbative uncertainty in QCD could be significantly reduced should the $\alpha_s^4$ corrections be available.
\item The softening of the singularities that is observed in \Lb leads to series that approach
their true value uniformly but somewhat slowly. The results for larger $n$ are further away from their
true value which would translate into larger values of $\alpha_s$ for higher $n$ in an extraction
of the strong coupling. In Refs.~\cite{Boito:2019pqp,Boito:2020lyp} this behaviour was found in the QCD analysis, and the partial renormalon cancellation that
we found in this work offers a plausible explanation for this trend.
\item We should also point out that in \Lb the charm pseudo-scalar and vector moment ratios behave rather similarly with respect to scale variations. This is different from what is observed in QCD, where results from the $P$ correlator tend to have
larger perturbative errors~\cite{Boito:2019pqp,Boito:2020lyp}. Therefore, it seems that what is causing this qualitative difference is beyond the $1/\beta_0$ approximation.
\item Finally, the results for the large-$\beta_0$ limit of the ratios of moments can be used to derive new combinations of $R_{q,n}^\delta$ guided by renormalon cancellations that optimize the behaviour of the perturbative series. Reducing the spread of scale variations at $\mathcal{O}(\alpha_s^3)$ could lead to significant reductions on the final error in $\alpha_s$ determinations based on heavy-quark current correlators.
\end{itemize}

\section{Conclusions}
\label{sec:Conclusions}

We have obtained the small-momentum expansion of the vector, axial-vector, scalar, and pseudo-scalar
correlators in the \Lb limit of QCD. The results
for the vector correlator for low values of $n$ were known since the work of
Ref.~\cite{Grozin:2004ez} while the others are new.

We have used these results to gain understanding about general features of the perturbative series for
the ratios of moments $R_{q,n}^\delta$ of Eq.~\eqref{eq:RqnDef}. Ratios $R_{c,n}^P$ have been used
since some time for the extraction of the strong coupling from lattice results for the pseudo-scalar
charm correlator, while ratios $R_{q,n}^V$ with $q=c,$ $b$ were recently shown to lead to competitive
and reliable extractions of $\alpha_s$ from data for \mbox{$\epem$} with charm or bottom flavour content.
We identified partial renormalon cancellations that make the series for these ratios of moments better behaved than the series of moments
$M_{q,n}^V$. These cancellations, however, is accompanied by a slower convergence towards the expected
results. This observation provides a plausible explanation for the tendency to larger values of $\alpha_s$ with
increasing $n$ observed in Refs.~\cite{Boito:2019pqp,Boito:2020lyp}, although in QCD the effect appears to be less pronounced than in \Lb.

Another observation that can be drawn from the \Lb results is that the series at $\mathcal{O}(\alpha_s^3)$ are still somewhat far from the expected values and still display a significant variation with renormalisation scale. This means that having the $\mathcal{O}(\alpha_s^4)$ term for $R_{q,n}^V$ would, very likely, significantly improve the $\alpha_s$ extractions from $R_{q,n}^{(V,P)}$ in terms of both central values and perturbative uncertainty. At present, we are aware of an ongoing calculation of the $\mathcal{O}(\alpha_s^4)$ correction for $M_{q,1}^V$~\cite{AMaierTalk}. To obtain the ratios at this order, however,
the results for higher $n$ would be required.

The renormalon structure of the ratios of moments obtained in the large-$\beta_0$ limit can also be used to design combinations of $R_{q,n}^\delta$ that display weaker scale variations and that could approach the expected value faster. Provided that these combinations can be reliably obtained from experimental and lattice data, they could be the basis for improved determinations of $\alpha_s$ from heavy-quark current correlators.

The results presented here can also have implications for the heavy-quark mass extractions from $\epem$
and from lattice data for the pseudo-scalar charm correlator. In the literature, the final perturbative uncertainty on the quark masses is estimated using different prescriptions
for the renormalisation scale variation. We intend to use the \Lb results presented here to shed light
on this aspect of the quark mass extractions.
A possibility to be explored is the construction of combinations of roots of moments $(M_n^\delta)^{1/2n}$, linearly sensitive to the quark mass, in the same spirit as the discussion in Sec.~\ref{sec:NewRatios}, aiming at partial renormalon cancellations and better perturbative behavior at $\mathcal{O}(\alpha_s^3)$, with the potential of improving the determinations of heavy-quark masses,
but this is left for future work.
The results we have obtained should also allow for a connection
with non-relativistic QCD, since we were able to obtain the small-momentum expansion of the correlators for large values of $n$, which leave the domain of the relativistic sum rules. The investigation of this connection is also beyond the scope of this work and should be explored in the future.

\subsection*{Acknowledgements}
This work was supported in part by the SPRINT project funded by the S\~ao Paulo Research Foundation (FAPESP) and the University of
Salamanca, grant No.\ \mbox{2018/14967-4}.
DB's work was supported by FAPESP Grant
No.\ 2015/20689-9, by CNPq Grant No.\ 309847/2018-4, and by the Coordena\c c\~ao de Aperfei\c coamento de Pessoal de N\'ivel Superior –- Brasil (CAPES) –- Finance Code 001. MVR is supported by FAPESP grant No.\ 2019/16957-9. VM is supported by
the MECD grant
PID2019-105439GB-C22, the IFT Centro de Excelencia Severo Ochoa Program under Grant SEV-2012-0249, the EU STRONG-2020 project under the program
H2020-INFRAIA-2018-1, grant agreement No.\ 824093 and the COST Action CA16201 PARTICLEFACE.

\appendix

\section{Small-momentum expansion of the two-loop integrals}
\label{app:ExpLoop}

In the calculation of the Feynman diagrams of Fig.~\ref{fig:two-loop},
after the trace is performed and the powers of momenta in the numerator are written in terms of the propagators, the problem is reduced
to the study of the following two-loop generic scalar integral
\begin{align}\label{eq:J2}
&J_2(q^2;n_1,..., n_5) = \nn \\
&= \int \frac{{\rm d}^d k_1\,{\rm d}^dk_2}{[(k_1+q)^2-m_q^2]^{n_1}[(k_2+q)^2-m_q^2]^{n_2}[k_1^2-m_q^2]^{n_3}[k_2^2-m_q^2]^{n_4}[(k_2-k_1)^2]^{n_5}}\,,
\end{align}
where $n_5$, the exponent of the gluon propagator, acts as an analytic regulator and therefore the integral must be carried out for a generic value of this parameter ($n_i$ with $1\leq i\leq 4$ are always integer numbers).
Because of this constraint, it is not possible to use integration-by-parts~\cite{Chetyrkin:1981qh} to reduce the problem to the calculation of a
small set of master integrals that can be expanded in $q^2$ using modern techniques such as the Mellin-Barnes transform~\cite{Friot:2005cu}.
Therefore, we perform the asymptotic small-momentum expansion by successive applications of the d'Alembertian operator in momentum space
\begin{equation}
\Box_q = \frac{\partial}{\partial q_\mu \partial q^\mu}\,.
\end{equation}
The corresponding Taylor expansion can be cast as~\cite{Davydychev:1992mt}:
\begin{equation}
J_2(q^2;n_1,..., n_5) = \sum_{j=0}^{\infty} \frac{1}{j! (d/2)_j} \biggl(\frac{q^2}{4} \biggr)^{\!j} \big[ \Box_q^j J_2(q^2;n_1,..., n_5) \big]_{q=0}\,,
\end{equation}
where $(a)_j \equiv \G(a+j)/\G(a)$ is the Pochhammer symbol. The application of the d'Alembertian operator on the integrals $J_2$ results in
\begin{align}
\Box_q J_2 = &\,4\big\{(n_1+n_2+1-d/2)[n_1\, \boldsymbol{1^+}J_2+n_2 \boldsymbol{2^+}J_2]\nn \\&
+m^2[n_1(n_1+1)\boldsymbol{1^{++}}J_2 + n_2(n_2+1)\boldsymbol{2^{++}}J_2] \nn \\
& + n_1 n_2 [2m^2 \boldsymbol{1^+}\boldsymbol{2^+}J_2 -\boldsymbol{1^+}\boldsymbol{2^+}\boldsymbol{5^-} J_2]\big\}\,,
\end{align}
where we used the notation $\boldsymbol{1^\pm}J_2(q^2;n_1,n_2,n_3,n_4,n_5) = J_2(q^2;n_1\pm1,n_2,n_3,n_4,n_5)$ and analogously for $\boldsymbol{2^\pm}$ and $\boldsymbol{5^\pm}$, with $\boldsymbol{n^{++}}\equiv (\boldsymbol{n^+})^2$. Higher derivatives are obtained by recursively applying the d'Alembertian operator. After setting $q^2=0$, the remaining tadpole single-scale integrals can be solved analytically~\cite{Smirnov:2004ym}
\begin{align}
J_2(0;n_1,\ldots,n_5)&= -\pi^d (-1)^{\lambda_1+\lambda_2+\lambda_3}(m_q^2)^{d-\lambda_1-\lambda_2-\lambda_3} \\
&\times\frac{\Gamma(\lambda_1+\lambda_3-d/2)\Gamma(\lambda_2+\lambda_3-d/2)\Gamma(d/2-\lambda_3)\Gamma(\lambda_1+\lambda_2+\lambda_3-d)}{\Gamma(\lambda_1)\Gamma
(\lambda_2)\Gamma(\lambda_1+\lambda_2+2\lambda_3-d)\Gamma(d/2)}\,,\nn
\end{align}
where $\lambda_1 \equiv n_1+n_3$, $\lambda_2 \equiv n_2+n_4$ and $\lambda_3 \equiv n_5$.

\section{Explicit results}
\label{app:ExplicitResults}

\subsection{Polynomials}
\label{app:Polynomials}

In this appendix we give explicitly the first three polynomials $P_n^\delta(u)$ for the four quark currents considered in this work. Results for higher values of $n$ for the vector and pseudo-scalar currents are available in~\cite{gitlabFile}.

Vector current:
\begin{align}\label{eq:Pvec}
P_1^V(u) &= 3+\frac{92 u}{27}+\frac{29 u^2}{27}+\frac{u^3}{9}\,, \\
P_2^V(u) &= 10+\frac{2095 u}{162}+\frac{7393 u^2}{1296}+\frac{2887 u^3}{2592}+\frac{7
u^4}{54}+\frac{u^5}{96}\,, \nn \\
P_3^V(u) &= \frac{315}{16}+\frac{54791 u}{1920}+\frac{62653 u^2}{3840}+\frac{3039 u^3}{640}+\frac{1037
u^4}{1280}+\frac{u^5}{10}+\frac{19 u^6}{1920}+\frac{u^7}{1920}\,.\nn
\end{align}

Pseudo-scalar current:
\begin{align}\label{eq:PP}
P_0^P(u) &=-\frac{2u}{3}(7+u)\,, \\
P_1^P(u) &= 6-\frac{11 u}{18}-\frac{49 u^2}{12}-\frac{8 u^3}{9}-\frac{u^4}{12}\,, \nn\\
P_2^P(u) &= \frac{2-u}{2}\biggl(\frac{u^5}{192}+\frac{19 u^4}{192}+\frac{467 u^3}{576}+\frac{2311 u^2}{576}+\frac{2677 u}{288}+\frac{15}{2}\biggr).\nn
\end{align}

Scalar current:
\begin{align}\label{eq:PS}
P_0^S(u) &= u\biggl(-\frac{61}{27} +\frac{235 u}{27}+\frac{260 u^2}{27}+\frac{20 u^3}{9}+\frac{2 u^4}{9} \biggr)\,, \\
P_1^S(u) &= 15+\frac{703 u}{36}+\frac{2333 u^2}{72}+\frac{2539 u^3}{72}+\frac{305 u^4}{18}+\frac{197 u^5}{48}+\frac{7 u^6}{12}+\frac{5 u^7}{144}\,, \nn\\
P_2^S(u) &= \frac{105}{2}+\frac{41357 u}{480}+\frac{15517 u^2}{160}+\frac{513613 u^3}{5760}+\frac{99889 u^4}{1920}+\frac{35993 u^5}{1920} \nn\\
&+\frac{1711 u^6}{384}+\frac{223 u^7}{320}+\frac{u^8}{16}+\frac{7 u^9}{2880}\,.\nn
\end{align}

Axial-vector current:
\begin{align}
P_1^A(u) &= 6+\frac{661 u}{54}+\frac{1423 u^2}{108}+\frac{271 u^3}{36}+\frac{205 u^4}{108}+\frac{7 u^5}{36}\,, \\
P_2^A(u) &= 30+\frac{2161 u}{36}+\frac{8315 u^2}{144}+\frac{30793 u^3}{864}+\frac{1555 u^4}{108}+\frac{98 u^5}{27}+\frac{25 u^6}{48}+\frac{u^7}{32}\,, \nn \\
P_3^A(u) &= \frac{315}{4}+\frac{77507 u}{480}+\frac{798 u^2}{5}+\frac{59687 u^3}{576}+\frac{337453 u^4}{6912}+\frac{580397 u^5}{34560} \nn \\
&+\frac{69961u^6}{17280}+\frac{10969 u^7}{17280}+\frac{1973 u^8}{34560}+\frac{77 u^9}{34560}\,.\nn
\end{align}

\subsection[Leading-\texorpdfstring{$n_\ell$}{nl} coefficients]{\boldmath Leading-\texorpdfstring{$n_\ell$}{nl} coefficients} \label{app:leading_nl}

Here we give the leading-$n_\ell$ coefficients in the perturbative expansion of $N_n^\delta\,C_n^\delta$, with $\mu = \overline{m}_q$, up to $\alpha_s^4$ for the first four physical moments of each correlator.
The coefficients of order $n_\ell^3 \alpha_s^4$ for $\delta = P,S,A$ are new in the literature. In the results of this section we define\footnote{Not to be confused with $a_\mu$ defined in Eq.~\eqref{eq:beta}.}
\begin{equation}
\tilde C_{n}^\delta \equiv N_n^\delta C_n^\delta\,,\qquad {\rm and}\qquad a_s\equiv\frac{\alpha_s}{\pi}.
\end{equation}

Vector correlator:
\begin{align}
\tilde C_1^V =\,& 1.0667\, +2.5547 \,a_s+ (\cdots+0.66228 \,n_\ell)a_s^2 \\
& + (\cdots+0.096101 \,n_\ell^2)a_s^3 +(\cdots+ 0.096093 \,n_\ell^3)a_s^4\,, \nn\\
\tilde C_2^V =\,& 0.45714\, +1.1096 \,a_s+ (\cdots+0.45492 \,n_\ell)a_s^2 \nn\\
& + (\cdots-0.01595 \,n_\ell^2)a_s^3 +(\cdots+ 0.036331 \,n_\ell^3)a_s^4 \,, \nn\\
\tilde C_3^V =\,& 0.27090\, +0.51940 \,a_s+ (\cdots+0.42886 \,n_\ell)a_s^2 \nn\\
&+ (\cdots-0.039596 \,n_\ell^2)a_s^3 +(\cdots +0.033047 \,n_\ell^3)a_s^4 \,, \nn\\
\tilde C_4^V =\,& 0.18471\, +0.20312 \,a_s+ (\cdots+0.42483 \,n_\ell)a_s^2 \nn\\
&+ (\cdots-0.052774 \,n_\ell^2)a_s^3
+(\cdots +0.033935 \,n_\ell^3)a_s^4\, .\nn
\end{align}

Pseudo-scalar correlator:
\begin{align}
\tilde C_0^P =\,& 1.3333\, +3.1111 \,a_s+ (\cdots+0.61729 \,n_\ell)a_s^2 \\
&+ (\cdots+0.37997 \,n_\ell^2)a_s^3 + (\cdots +0.22899 \,n_\ell^3) a_s^4\,,\nn \\
\tilde C_1^P =\,& 0.53333\, +2.0642 \,a_s+ (\cdots+0.28971 \,n_\ell)a_s^2 \nn\\
& + (\cdots+0.070202 \,n_\ell^2)a_s^3+ (\cdots +0.035807 \,n_\ell^3) \,a_s^4\,, \nn\\
\tilde C_2^P =\,& 0.30477\, +1.2117 \,a_s+ (\cdots+0.26782 \,n_\ell)a_s^2 \nn\\
& + (\cdots+0.015357 \,n_\ell^2)a_s^3+ (\cdots +0.021840 \,n_\ell^3)a_s^4\,, \nn\\
\tilde C_3^P =\,& 0.20318 \, + 0.71276 \,a_s+ (\cdots+0.28628 \,n_\ell)a_s^2\nn \\
& + (\cdots -0.0091663 \,n_\ell^2)a_s^3
+ (\cdots +0.021261 \,n_\ell^3) \,a_s^4\,.\nn
\end{align}

Scalar correlator:
\begin{align}
\tilde C_0^S =\,& 0.8\, +0.60247 \,a_s+ (\cdots+0.58765 \,n_\ell)a_s^2 \\
&+ (\cdots+0.23981 \,n_\ell^2)a_s^3+(\cdots +0.20536 \,n_\ell^3) a_s^4\,, \nn\\
\tilde C_1^S =\,& 0.22857\, +0.42582 \,a_s+ (\cdots+0.23664 \,n_\ell)a_s^2\nn\\
&+ (\cdots+0.0039812 \,n_\ell^2)a_s^3+(\cdots+0.030916 \,n_\ell^3) a_s^4\,,\nn \\
\tilde C_2^S =\,& 0.10159\, +0.15356 \,a_s+ (\cdots+0.15634 \,n_\ell)a_s^2 \nn\\
& + (\cdots-0.018026 \,n_\ell^2)a_s^3+(\cdots +0.017163 \,n_\ell^3) a_s^4\,,\nn \\
\tilde C_3^S =\,& 0.055411\, + 0.032800 \,a_s+ (\cdots+0.12383 \,n_\ell)a_s^2\nn \\
& + (\cdots-0.020909 \,n_\ell^2)a_s^3
+(\cdots +0.013605 \,n_\ell^3) \,a_s^4\,.\nn
\end{align}

Axial-vector correlator:
\begin{align}
\tilde C_1^A =\,& 0.53333\, +0.84609 \,a_s+ (\cdots+0.41317 \,n_\ell)a_s^2 \\
& + (\cdots+0.047848 \,n_\ell^2)a_s^3+(\cdots+0.069840 \,n_\ell^3)a_s^4\,,\nn \\
\tilde C_2^A =\,& 0.15238\, +0.14166 \,a_s+ (\cdots+0.19218 \,n_\ell)a_s^2 \nn\\
& + (\cdots-0.020498 \,n_\ell^2)a_s^3+(\cdots +0.017170 \,n_\ell^3)a_s^4\,,\nn \\
\tilde C_3^A =\,& 0.067725\, -0.012760 \,a_s+ (\cdots+0.13562 \,n_\ell)a_s^2 \nn\\
& + (\cdots-0.022336 \,n_\ell^2)a_s^3+(\cdots +0.012418 \,n_\ell^3)a_s^4\,,\nn \\
\tilde C_4^A =\,& 0.036941\, -0.057469 \,a_s+ (\cdots+0.10678 \,n_\ell)a_s^2\nn \\
& + (\cdots-0.020499 \,n_\ell^2)a_s^3
+(\cdots +0.010501 \,n_\ell^3)a_s^4\,.\nn
\end{align}

\bibliographystyle{jhep}
\bibliography{BibTex_Ratios_Largeb0}

\end{document}